\documentclass[a4paper]{article}%
\usepackage{amsmath}
\usepackage{amsfonts}
\usepackage{amssymb}
\usepackage{graphicx}%
\usepackage{multirow}
\usepackage{subcaption}
\providecommand{\U}[1]{\protect\rule{.1in}{.1in}}
\newtheorem{theorem}{Theorem}

\newtheorem{definition}[theorem]{Definition}

\newtheorem{remark}[theorem]{Remark}

\begin{document}
\title{\textbf{Robust Estimators and Test-Statistics for One-Shot Device Testing
Under the Exponential Distribution}}
\author{N. Balakrishnan, E. Castilla, N. Mart\'in, and L.Pardo}
\maketitle
\begin{abstract}
This paper develops a new family of estimators, the minimum density power
divergence estimators (MDPDEs), for the parameters of the one-shot device
model as well as a new family of test statistics, Z-type test statistics based
on MDPDEs, for testing the corresponding model parameters. The
family of MDPDEs contains as a particular case the maximum likelihood
estimator (MLE) considered in Balakrishnan and Ling (2012). Through a
simulation study, it is shown that some MDPDEs have a better behavior than the
MLE\ in relation to  robustness. At the same time, it can be seen that some
Z-type tests based on MDPDEs have a better behavior than the classical Z-test
statistic also in terms of robustness.
\end{abstract}
\section{Introduction\label{sec1}}

The reliability of a product, system, weapon, or piece of equipment can be defined as the ability of the device to perform as designed, or, more simply, as the probability that the device does not fail when used. Engineers's assess reliability by repeatedly testing the device and observing its failure rate. Certain products, called \textquotedblleft one-shot\textquotedblright \ devices, make this approach challenging. One-shot devices can only be used once and after use the device is either destroyed or must be rebuilt. Consequently, one can only know whether the failure time is either before or after the test time. The outcomes from each of the devices are therefore binary, either left-censored (failure) or right-censored (success). Some examples of one-shot devices are nuclear weapons, space shuttles, automobile air bags, fuel injectors, disposable napkins, heat detectors, and fuses. In survival analysis, these data are called \textquotedblleft current status data\textquotedblright. For instance, in animal carcinogenicity experiments, one observes whether a tumor occurs at the examination time for each subject.

Due to the advances in manufacturing design and technology, products have now become highly reliable with long lifetimes. This fact would pose a problem in the analysis if only few or no failures are observed. For this reason, accelerated life tests are often used by adjusting a controllable factor such as temperature in order to have more failures in the experiment. On the other hand, accelerated life testing would shorten the experimental time and also help to reduce the experimental cost. In this paper, we shall assume that the failure times of  devices follow an exponential distribution. In this context, Balakrishnan and Ling (2012) developed the EM algorithm for finding the maximum likelihood estimators of the model parameters. Fan et al. (2009) studied a Bayesian approach for one-shot device testing along with an accelerating factor, in which the failure times of  devices is assumed to follows once again an exponential distribution. Rodrigues et al. (1993) presented two approaches based on the likelihood ratio statistics and the posterior Bayes factor for comparing several exponential accelerated life models. Chimitova and Balakrishnan (2015) made a comparison of several goodness-of-fit tests for one-shot device testing.

In Section \ref{sec2}, we  present a description of the one-shot device model as well as the maximum likelihood estimators for  the model parameters. Section \ref{sec3} develops the minimum density power divergence estimator as a natural extension of the maximum likelihood estimator, as well as its asymptotic distribution. In Section \ref{sec4}, $Z$-type test statistics are introduced in order to test some hypotheses about the parameters of the one-shot device model. Some numerical examples are presented in Section \ref{sec5}, with one of them relating to a reliability situation and the other two are real applications to tumorigenicity experiments. In Section \ref{sec6}, an extensive simulation study is presented in order to analyze the robustness of the MDPDEs, as well as the $Z$-type test introduced earlier. Finally, some concluding remarks are made in Section \ref{sec7}.

\section{Model formulation and maximum likelihood estimator\label{sec2}}

Consider a reliability testing experiment in which at each time, $t_{j}$, $j=1,2,...,J$, $K$ devices are placed in total under temperatures $w_{i}$, $i=1,...,I$. Therefore, $IJK$ devices are tested in total at temperatures $w_{i}$, $i=1,...,I$, at times $t_{j},$ $j=1,...,J$. It is worth noting that a successful detonation occurs if the lifetime is beyond the inspection time, whereas the lifetime will be before the inspection time if the detonation is a failure. For each temperature $w_{i}$ and at each inspection time $t_{j}$, the number of failures, $n_{ij}$, is then recorded.

In Balakrishnan and Ling (2012), an example is illustrated, in which $30$ devices were tested at temperatures $w_{i}\in\{35,$ $45,$ $55\}$, each with $10$ units being detonated at times $t_{j}\in\{10,$ $20,$ $30\}$, respectively.$\ $In this example, we have $I=3$, $J=3$ and $K=10$. The number of failures observed is summarized in the $3\times3$ table given in Table \ref{table1}. In this one-shot device testing experiment, there were in all $48$ failures out a total of $90$ tested devices.%

\begin{table}[!h]
\renewcommand{\arraystretch}{1.3}
\caption{Failures in the-shot device testing experiment of Balakrishnan and Ling (2012).\label{table1}}%
\centering
$%
\begin{tabular}
[c]{l|l|l|l|}\cline{2-4}
& $t_{1}=10$ & $t_{2}=20$ & $t_{3}=30$\\\hline
\multicolumn{1}{|l|}{$w_{1}=35$} & \multicolumn{1}{|c|}{$3$} &
\multicolumn{1}{|c|}{$3$} & \multicolumn{1}{|c|}{$7$}\\\hline
\multicolumn{1}{|l|}{$w_{2}=45$} & \multicolumn{1}{|c|}{$1$} &
\multicolumn{1}{|c|}{$5$} & \multicolumn{1}{|c|}{$7$}\\\hline
\multicolumn{1}{|l|}{$w_{3}=55$} & \multicolumn{1}{|c|}{$6$} &
\multicolumn{1}{|c|}{$7$} & \multicolumn{1}{|c|}{$9$}\\\hline
\end{tabular}
\ \ \ \ \ $%
\end{table}

We shall assume here, in accordance  Balakrishnan and Ling$\ $(2012), that the true lifetimes $T_{ijk}$, where $i=1,2,...,I$, $j=1,2,...,J$, $k=1,...,K$, are independent and identically distributed exponential random variables with probability density function
\[
f(t|\lambda)=\lambda\exp\left(  -\lambda t\right)  ,
\]
where $\lambda>0$ is the unknown failure rate. In practice, we consider inspection times $t_{j}$, $j=1,...,J$, rather than $t>0$, and we relate the parameter $\lambda$ to an accelerating factor of temperature $w_{i}>0$ through a log-linear link function as
\[
\lambda_{w_{i}}(\boldsymbol{\alpha})=\alpha_{0}\exp\left\{  \alpha_{1}%
w_{i}\right\}  ,
\]
where $\alpha_{0}>0$ and $\alpha_{1}\in\mathbb{R}$ are unknown parameters. Therefore, the corresponding distribution function is%
\begin{align}
F(t_{j}|\lambda_{w_{i}}(\boldsymbol{\alpha}))&=1-\exp\left\{  -\lambda_{w_{i}%
}(\boldsymbol{\alpha})t_{j}\right\}  \nonumber \\
&  =1-\exp\left\{  -\alpha_{0}\exp\left\{
\alpha_{1}w_{i}\right\}  t_{j}\right\}  \label{eq:expo}%
\end{align}
and the density function%
\begin{equation}
f(t_{j}|\lambda_{w_{i}}(\boldsymbol{\alpha}))=\alpha_{0}\exp\left\{
\alpha_{1}w_{i}\right\}  \exp\left\{  -\alpha_{0}\exp\left\{  \alpha_{1}%
w_{i}\right\}  t_{j}\right\}  . \label{eq:expo2}%
\end{equation}
The data are completely described on $K$ devices, through the contingency table of failures $\boldsymbol{n}=(n_{11},...,n_{1J},...,\allowbreak
n_{I1},...,n_{IJ})^{T}$, collected at the temperatures $\boldsymbol{w}%
=(w_{1},...,w_{I})^{T}$ and the inspection times $\boldsymbol{t}%
=(t_{1},...,t_{J})^{T}$.

We shall consider the theoretical probability vector $\boldsymbol{p}%
(\boldsymbol{\alpha})$ defined by
\begin{align*}
\boldsymbol{p}(\boldsymbol{\alpha})=&\left(  \tfrac{F(t_{1}|\lambda_{w_{1}%
}(\boldsymbol{\alpha}))}{IJ},\tfrac{1-F(t_{1}|\lambda_{w_{1}}%
(\boldsymbol{\alpha}))}{IJ}, \right. \\
&\left....,\tfrac{F(t_{J}|\lambda_{w_{I}}%
(\boldsymbol{\alpha}))}{IJ},\tfrac{1-F(t_{J}|\lambda_{w_{I}}%
(\boldsymbol{\alpha}))}{IJ}\right)  ^{T},
\end{align*}
as well as the observed probability vector%
\[
\widehat{\boldsymbol{p}}=\left(  \tfrac{n_{11}}{IJK},\tfrac{K-n_{11}}%
{IJK},...,\tfrac{n_{IJ}}{IJK},\tfrac{K-n_{IJ}}{IJK}\right)  ^{T},
\]
both of dimension $2IJ$. Then the Kullback-Leibler divergence between the probability vectors $\widehat{\boldsymbol{p}}$\ and $\boldsymbol{p}%
(\boldsymbol{\alpha})$ is given by
\begin{align*}
d_{KL}\left(  \widehat{\boldsymbol{p}},\boldsymbol{p}(\boldsymbol{\alpha
})\right)  &=\frac{1}{IJ}\sum\limits_{i=1}^{I}\sum\limits_{j=1}^{J}\left(
\frac{n_{ij}}{K}\log\frac{n_{ij}}{KF(t_{j}|\lambda_{w_{i}}(\boldsymbol{\alpha
}))}\right.\\
&+\left. \frac{K-n_{ij}}{K}\log\frac{K-n_{ij}}{K\left(  1-F(t_{j}|\lambda_{w_{i}%
}(\boldsymbol{\alpha}))\right)  }\right) .
\end{align*}
It is not difficult to establish the following result.

\vspace{3pt}
 \begin{theorem}
 The likelihood function%
\begin{align*}
\mathcal{L}\left(  \boldsymbol{\alpha}\left\vert K,\boldsymbol{n}%
,\boldsymbol{t},\boldsymbol{w}\right.  \right)  =\prod\limits_{i=1}^{I}%
\prod\limits_{j=1}^{J}F(t_{j}|\lambda_{w_{i}}(\boldsymbol{\alpha}))^{n_{ij}}\\
\left(  1-F(t_{j}|\lambda_{w_{i}}(\boldsymbol{\alpha}))\right)  ^{K-n_{ij}},
\end{align*}
where $F(t_{j}|\lambda_{w_{i}}(\boldsymbol{\alpha}))$ is given by
(\ref{eq:expo}), is related to the Kullback-Leibler divergence between the
probability vectors $\widehat{\boldsymbol{p}}$\ and $\boldsymbol{p}%
(\boldsymbol{\alpha})$ through%
\begin{align}
d_{KL}\left(  \widehat{\boldsymbol{p}},\boldsymbol{p}(\boldsymbol{\alpha
})\right)  =\frac{1}{IJK}\left(  s-\log\mathcal{L}\left(  \boldsymbol{\alpha
}\left\vert K,\boldsymbol{n},\boldsymbol{t},\boldsymbol{w}\right.  \right)
\right)  , \label{1}%
\end{align}
with $s$ being a constant not dependent on $\boldsymbol{\alpha}$.
\end{theorem}
\vspace{3pt}

Based on the previous result, we have the following definition for the maximum
likelihood estimators of $\alpha_{0}$ and $\alpha_{1}.$

\vspace{3pt}
\begin{definition}
We consider the data given by $K$, $\boldsymbol{n}$, $\boldsymbol{t}$,
$\boldsymbol{w}$ for the one-shot device model. Then, the maximum likelihood
estimator of $\boldsymbol{\alpha}=(\alpha_{0},\alpha_{1})^{T}$,
$\widehat{\boldsymbol{\alpha}}=(\widehat{\alpha}_{0},\widehat{\alpha}_{1}%
)^{T}$, can be defined as
\begin{equation}
\widehat{\boldsymbol{\alpha}}=\arg\min_{\boldsymbol{\alpha}\in\Theta}%
d_{KL}\left(  \widehat{\boldsymbol{p}},\boldsymbol{p}(\boldsymbol{\alpha
})\right)  , \label{2}%
\end{equation}
where $\Theta=(\mathbb{R}^{+},\mathbb{R})^{T}$.
\end{definition}
\vspace{3pt}

\section{Minimum density power divergence estimator\label{sec3}}

Based on expression (\ref{2}), we can think of defining  an estimator minimizing any distance or divergence between the probability vectors $\widehat{\boldsymbol{p}}$ and $\boldsymbol{p}(\boldsymbol{\alpha})$. There are  many different divergence measures (or distances) known in the lierature, see, for instance, Pardo (2006) and Basu et al. (2011), and  the natural question is if all of them are valid to define estimators with good properties. Initially the answer is yes,\ but we must think in terms of efficiency as well as  robustness of the defined estimators. From an asymptotic point of view, it is well-known that the maximum likelihood estimator is a BAN (Best
Asymptotically Normal) estimator, but at the same time we know that the maximum likelihood estimator has a very poor behavior, in general, in relation to  robustness. It is well-known that a gain in robustness leads to a loss of efficiency. Therefore, the distances (divergence measures) that we must use are those which result in estimators having good properties in terms of robustness with low loss of efficiency. The density power divergence measure introduced by Basu et al. (1998) has the required properties and has been studied for many different problems until now. For more details, see Ghosh et al. (2016), Basu et al. (2016) and the references therein.

Based on Ghosh and Basu (2013), the MDPDE of $\boldsymbol{\alpha}$ is first introduced, and later in Result \ref{Th4} it is shown that this estimator can be considered as a natural extension of (\ref{2}).

\vspace{3pt}
\begin{definition}
\label{def1}Let $y_{ijk}$ , $i=1,2,...,I$, $j=1,2,...,J$, $k=1,...,K$, be a
sequence of independent Bernoulli random variables, $y_{ijk}\overset{ind}{\sim
}Ber(\pi_{ij}(\boldsymbol{\alpha}))$, such that $\pi_{ij}(\boldsymbol{\alpha
})=F\left(  t_{j}|\lambda_{w_{i}}(\boldsymbol{\alpha})\right)  $ and $n_{ij}=%
{\textstyle\sum\nolimits_{k=1}^{K}}
y_{ijk}$. The MDPDE of $\boldsymbol{\alpha}$, with tuning parameter $\beta
\geq0$, is given by
\begin{equation}
\widehat{\boldsymbol{\alpha}}_{\beta}=\arg\min_{\boldsymbol{\alpha}\in\Theta
}\frac{1}{IJK}\sum\limits_{i=1}^{I}\sum\limits_{j=1}^{J}\sum\limits_{k=1}%
^{K}V_{ij}\left(  y_{ijk},\beta\right),  \label{4}%
\end{equation}
where
\begin{align*}
V_{ij}\left(  y_{ijk},\beta\right)  &=\pi_{ij}^{\beta+1}(\boldsymbol{\alpha })+(1-\pi_{ij}(\boldsymbol{\alpha}))^{\beta+1}\\
&-\frac{1+\beta}{\beta}\left(\pi_{ij}^{y_{ijk}}(\boldsymbol{\alpha})(1-\pi_{ij}(\boldsymbol{\alpha }))^{1-y_{ijk}}\right)  ^{\beta}.
\end{align*}
\end{definition}
\vspace{3pt}

For more details about the interpretation of Definition 3, see formula 2.3 in
Ghosh and Basu (2013), in which $\pi_{ij}^{y_{ijk}}(\boldsymbol{\alpha}%
)(1-\pi_{ij}(\boldsymbol{\alpha}))^{1-y_{ijk}}$ plays the role of the density
in our context. Notice that the expression to be minimized in (\ref{4}) can be
simplified as
\begin{align}
&  \frac{1}{IJK}\sum\limits_{i=1}^{I}\sum\limits_{j=1}^{J}\sum\limits_{k=1}%
^{K}\left\{  \pi_{ij}^{\beta+1}(\boldsymbol{\alpha})+(1-\pi_{ij}%
(\boldsymbol{\alpha}))^{\beta+1}\right. \nonumber \\
&\left. -\frac{1+\beta}{\beta}\left(  \pi_{ij}^{y_{ijk}}(\boldsymbol{\alpha})(1-\pi_{ij}(\boldsymbol{\alpha
}))^{1-y_{ijk}}\right)  ^{\beta}\right\} \nonumber\\
&  =\frac{1}{IJ}\sum\limits_{i=1}^{I}\sum\limits_{j=1}^{J}\left\{  \pi
_{ij}^{\beta+1}(\boldsymbol{\alpha})+(1-\pi_{ij}(\boldsymbol{\alpha}%
))^{\beta+1}\right. \nonumber \\
&\left. -\frac{1+\beta}{\beta}\frac{n_{ij}}{K}\pi_{ij}^{\beta
}(\boldsymbol{\alpha})-\frac{1+\beta}{\beta}\frac{K-n_{ij}}{K}(1-\pi
_{ij}(\boldsymbol{\alpha}))^{\beta}\right\}  . \label{4b}%
\end{align}

The following result provides an alternative expression for
$\widehat{\boldsymbol{\alpha}}_{\beta}$, given in Definition \ref{def1}%
,\ which is closer to (\ref{2}) in its expression, since only a divergence
measure between two probabilities is involved. Given two probability vectors
$\boldsymbol{p}=\left(  p_{1},...,p_{M}\right)  ^{T}$ and $\boldsymbol{q}%
=\left(  q_{1},...,q_{M}\right)  ^{T}$, the power density divergence measure
between $\boldsymbol{p}$ and $\boldsymbol{q}$, with tuning parameter $\beta
>0$, is given by%
\[
d_{\beta}\left(  \boldsymbol{p},\boldsymbol{q}\right)  =\sum\limits_{j=1}%
^{M}\left\{  q_{j}^{\beta+1}-(1+\tfrac{1}{\beta})q_{j}^{\beta}p_{j}+\tfrac
{1}{\beta}p_{j}^{1+\beta}\right\}  ,
\]
and for $\beta=0$,%
\[
d_{0}\left(  \boldsymbol{p},\boldsymbol{q}\right)  =\lim_{\beta\rightarrow
0^{+}}d_{\beta}\left(  \boldsymbol{p},\boldsymbol{q}\right)  =d_{KL}\left(
\boldsymbol{p},\boldsymbol{q}\right)  .
\]
Therefore, the density power divergence measure between the probability
vectors $\widehat{\boldsymbol{p}}$ and $\boldsymbol{p}(\boldsymbol{\alpha})$,
with tuning parameter $\beta>0$, has the  expression
\begin{align}
d_{\beta}\left(  \widehat{\boldsymbol{p}},\boldsymbol{p}(\boldsymbol{\alpha
})\right)   &  =\frac{1}{(IJ)^{\beta+1}}\sum\limits_{i=1}^{I}\sum
\limits_{j=1}^{J}\left\{  \pi_{ij}^{1+\beta}(\boldsymbol{\alpha})\right. \nonumber\\
&\left.-\tfrac{\beta+1}{\beta}\pi_{ij}^{\beta}(\boldsymbol{\alpha})\frac{n_{ij}}{K}%
+\tfrac{1}{\beta}\left(  \frac{n_{ij}}{K}\right)  ^{1+\beta}\right.
\nonumber\\
&  \left.  +\left(  1-\pi_{ij}(\boldsymbol{\alpha})\right)  ^{1+\beta}%
-\tfrac{\beta+1}{\beta}\left(  1-\pi_{ij}(\boldsymbol{\alpha})\right)
^{\beta}\tfrac{K-n_{ij}}{K} \right. \nonumber\\
&\left.+\tfrac{1}{\beta}\left(  \frac{K-n_{ij}}{K}\right)
^{1+\beta}\right\} , \label{3}%
\end{align}
and for $\beta=0$
\[
d_{\beta=0}\left(  \widehat{\boldsymbol{p}},\boldsymbol{p}(\boldsymbol{\alpha
})\right)  =\lim_{\beta\rightarrow0^{+}}d_{\beta}\left(
\widehat{\boldsymbol{p}},\boldsymbol{p}(\boldsymbol{\alpha})\right)
=d_{KL}\left(  \widehat{\boldsymbol{p}},\boldsymbol{p}(\boldsymbol{\alpha
})\right)  .
\]

\vspace{3pt}
\begin{theorem}
 \label{Th4}The MDPDE of $\boldsymbol{\alpha}$, with tuning parameter $\beta\geq0$, given in Definition \ref{def1}, can be alternatively defined as
\begin{equation}
\widehat{\boldsymbol{\alpha}}_{\beta}=\arg\min_{\boldsymbol{\alpha}\in\Theta
}d_{\beta}\left(  \widehat{\boldsymbol{p}},\boldsymbol{p}(\boldsymbol{\alpha
})\right)  , \label{3b}%
\end{equation}
where $d_{\beta}\left(  \widehat{\boldsymbol{p}},\boldsymbol{p}(\boldsymbol{\alpha})\right)  $ is as in (\ref{3}).\\
\end{theorem}
\vspace{3pt}

In the following result, the estimating equations needed to get the MDPDEs
are presented.

\vspace{3pt}
\begin{theorem}
\label{Th3}The MDPDE of $\boldsymbol{\alpha}$ with tuning parameter $\beta
\geq0$, $\widehat{\boldsymbol{\alpha}}_{\beta}$, can be obtained as the
solution of equations (\ref{5}) and (\ref{5B}).%
\vspace{3pt}

\begin{equation}
\sum\limits_{i=1}^{I}\sum\limits_{j=1}^{J}\left(  K\ F(t_{j}|\lambda_{w_{i}%
}(\boldsymbol{\alpha}))-n_{ij}\right)  f(t_{j}|\lambda_{w_{i}}%
(\boldsymbol{\alpha}))t_{j}\left[  F^{\beta-1}(t_{j}|\lambda_{w_{i}%
}(\boldsymbol{\alpha}))+\left(  1-F(t_{j}|\lambda_{w_{i}}(\boldsymbol{\alpha
}))\right)  ^{\beta-1}\right]  =0 \label{5}%
\end{equation}
\begin{equation}
\sum\limits_{i=1}^{I}\sum\limits_{j=1}^{J}\left(  K\ F(t_{j}|\lambda_{w_{i}%
}(\boldsymbol{\alpha}))-n_{ij}\right)  f(t_{j}|\lambda_{w_{i}}%
(\boldsymbol{\alpha}))t_{j}w_{i}\left[  F^{\beta-1}(t_{j}|\lambda_{w_{i}%
}(\boldsymbol{\alpha}))+\left(  1-F(t_{j}|\lambda_{w_{i}}(\boldsymbol{\alpha
}))\right)  ^{\beta-1}\right]  =0. \label{5B}%
\end{equation}

\end{theorem}

In the following results, the asymptotic distribution of the MDPDE of
$\boldsymbol{\alpha}$, $\widehat{\boldsymbol{\alpha}}_{\beta}$, for the
one-shot device model is presented.

\vspace{3pt}
\begin{theorem}
\label{Th5}The asymptotic distribution of the MDPDE
$\widehat{\boldsymbol{\alpha}}_{\beta}$ is given by
\[
\sqrt{K}\left(  \widehat{\boldsymbol{\alpha}}_{\beta}-\boldsymbol{\alpha}%
_{0}\right)  \overset{\mathcal{L}}{\underset{K\mathcal{\rightarrow}%
\infty}{\longrightarrow}}\mathcal{N}\left(  \boldsymbol{0},\boldsymbol{\bar
{J}}_{\beta}^{-1}(\boldsymbol{\alpha}_{0})\boldsymbol{\bar{K}}_{\beta
}(\boldsymbol{\alpha}_{0})\boldsymbol{\bar{J}}_{\beta}^{-1}(\boldsymbol{\alpha
}_{0})\right)  ,
\]
where
\begin{align}
\boldsymbol{\bar{J}}_{\beta}(\boldsymbol{\alpha})  &  =\sum\limits_{i=1}^{I}%
\begin{pmatrix}
\frac{1}{\alpha_{0}^{2}} & \frac{w_{i}}{\alpha_{0}} \nonumber \\
\frac{w_{i}}{\alpha_{0}} & w_{i}^{2}%
\end{pmatrix}
\sum\limits_{j=1}^{J}t_{j}^{2}f^{2}(t_{j}|\lambda_{w_{i}}(\boldsymbol{\alpha
}))\\
&\times \left[  F^{\beta-1}(t_{j}|\lambda_{w_{i}}(\boldsymbol{\alpha}%
))+(1-F(t_{j}|\lambda_{w_{i}}(\boldsymbol{\alpha})))^{\beta-1}\right]
,\label{Jbar}\\
\boldsymbol{\bar{K}}_{\beta}(\boldsymbol{\alpha})  &  =\sum\limits_{i=1}^{I}%
\begin{pmatrix}
\frac{1}{\alpha_{0}^{2}} & \frac{w_{i}}{\alpha_{0}}\\
\frac{w_{i}}{\alpha_{0}} & w_{i}^{2}%
\end{pmatrix}
\sum\limits_{j=1}^{J}t_{j}^{2}f^{2}(t_{j}|\lambda_{w_{i}}(\boldsymbol{\alpha
}))\nonumber\\
&  \times\left\{  \left[  F^{2\beta-1}(t_{j}|\lambda_{w_{i}}%
(\boldsymbol{\alpha}))+(1-F(t_{j}|\lambda_{w_{i}}(\boldsymbol{\alpha
})))^{2\beta-1}\right] \right.\nonumber \\
& \left. -\left[  F^{\beta}(t_{j}|\lambda_{w_{i}}%
(\boldsymbol{\alpha}))-(1-F(t_{j}|\lambda_{w_{i}}(\boldsymbol{\alpha
})))^{\beta}\right]  ^{2}\right\},  \label{Kbar}%
\end{align}
and $F(t_{j}|\lambda_{w_{i}}(\boldsymbol{\alpha}))$ and $f(t_{j}|\lambda_{w_{i}%
}(\boldsymbol{\alpha}))$ are given by (\ref{eq:expo}) and (\ref{eq:expo2}), respectively.
\end{theorem}
\vspace{3pt}

Since $\widehat{\boldsymbol{\alpha}}_{\beta=0}$ is the MLE of
$\boldsymbol{\alpha}$, obtained by maximizing $\log\mathcal{L}\left(
\boldsymbol{\alpha}\left\vert K,\boldsymbol{n},\boldsymbol{t},\boldsymbol{w}%
\right.  \right)$, or equivalently by minimizing%
\begin{align*}
d_{\beta=0}\left(  \widehat{\boldsymbol{p}},\boldsymbol{p}(\boldsymbol{\alpha
})\right)  &=\lim_{\beta\rightarrow0^{-}}d_{\beta}\left(
\widehat{\boldsymbol{p}},\boldsymbol{p}(\boldsymbol{\alpha})\right)
=d_{KL}\left(  \widehat{\boldsymbol{p}},\boldsymbol{p}(\boldsymbol{\alpha
})\right)  \\
&=\frac{1}{IJK}\left(  s-\log\mathcal{L}\left(  \boldsymbol{\alpha
}\left\vert K,\boldsymbol{n},\boldsymbol{t},\boldsymbol{w}\right.  \right)
\right)  ,
\end{align*}
the following result relates the asymptotic distribution of
$\widehat{\boldsymbol{\alpha}}_{\beta=0}$ given previously in terms of
$\boldsymbol{\bar{J}}_{\beta=0}(\boldsymbol{\alpha}_{0})$ and
$\boldsymbol{\bar{K}}_{\beta=0}(\boldsymbol{\alpha}_{0})$, with respect to the
Fisher information matrix, well-known in the classical asymptotic theory of
the MLEs.

\vspace{3pt}
\begin{theorem}
The asymptotic distribution of the MLE of $\boldsymbol{\alpha}$,
$\widehat{\boldsymbol{\alpha}}_{\beta=0}$, is%
\[
\sqrt{K}\left(  \widehat{\boldsymbol{\alpha}}_{\beta}-\boldsymbol{\alpha}%
_{0}\right)  \overset{\mathcal{L}}{\underset{K\mathcal{\rightarrow}%
\infty}{\longrightarrow}}\mathcal{N}\left(  \boldsymbol{0},\tfrac{1}%
{IJ}\boldsymbol{I}_{F}^{-1}\left(  \boldsymbol{\alpha}_{0}\right)  \right)  ,
\]
where%
\begin{footnotesize}
$$
\boldsymbol{I}_{F}\left(  \boldsymbol{\alpha}\right)  =\frac{1}{IJ}%
\sum\limits_{i=1}^{I}%
\begin{pmatrix}
\frac{1}{\alpha_{0}^{2}} & \frac{w_{i}}{\alpha_{0}}\\
\frac{w_{i}}{\alpha_{0}} & w_{i}^{2}%
\end{pmatrix}
\sum\limits_{j=1}^{J}t_{j}^{2}\frac{f^{2}(t_{j}|\lambda_{w_{i}}%
(\boldsymbol{\alpha}))}{F(t_{j}|\lambda_{w_{i}}(\boldsymbol{\alpha
}))(1-F(t_{j}|\lambda_{w_{i}}(\boldsymbol{\alpha})))}%
$$
\end{footnotesize}
\flushleft is the Fisher Information matrix for the one-shot device model. In addition,
relating the theory of MDPDEs with the Fisher Information matrix, we have
\[
\boldsymbol{J}_{\beta=0}(\boldsymbol{\alpha})=\boldsymbol{K}_{\beta
=0}(\boldsymbol{\alpha})=\boldsymbol{I}_{F}(\boldsymbol{\alpha}).
\]
\end{theorem}
\vspace{3pt}

\section{Robust Z-type tests\label{sec4}}

For testing the null hypothesis of a linear combination of $\boldsymbol{\alpha
}=(\alpha_{0},\alpha_{1})^{T}$, $H_{0}$: $m_{0}\alpha_{0}+m_{1}\alpha_{1}=d$,
or equivalently%
\begin{equation}
H_{0}\text{: }\boldsymbol{m}^{T}\boldsymbol{\alpha}=d, \label{W1}%
\end{equation}
where $\boldsymbol{m}^{T}=(m_{0},m_{1})$, it is important to know the asymptotic distribution of the MDPDE of $\boldsymbol{\alpha}$. In particular, in case we wish to test if the different temperatures do not affect the model of the one-shot devices, $\boldsymbol{m}^{T}=(m_{0},m_{1})=(0,1)$ and $d=0$ must  be considered. In the following definition, we present  $Z$-type test statistics based on $\widehat{\boldsymbol{\alpha}}_{\beta}$. Since  $Z$-type test statistics are a particular case of the Wald-type test, we can say that this type of robust test statistics have been considered previously in Basu et al. (2016) and Ghosh et al. (2016).

\vspace{3pt}
\begin{definition}
Let $\widehat{\boldsymbol{\alpha}}_{\beta}=(\widehat{\alpha}_{0,\beta},\widehat{\alpha}_{1,\beta})^{T}$ be the MDPDE of $\boldsymbol{\alpha}=(\alpha_{0},\alpha_{1})^{T}$. The family of $Z$-type test statistics for testing (\ref{W1}) is given by
\begin{equation}
Z_{K}(\widehat{\boldsymbol{\alpha}}_{\beta})=\sqrt{\frac{K}{\boldsymbol{m}%
^{T}\boldsymbol{\bar{J}}_{\beta}^{-1}(\widehat{\boldsymbol{\alpha}}_{\beta
})\boldsymbol{\bar{K}}_{\beta}(\widehat{\boldsymbol{\alpha}}_{\beta
})\boldsymbol{\bar{J}}_{\beta}^{-1}(\widehat{\boldsymbol{\alpha}}_{\beta
})\boldsymbol{m}}}(\boldsymbol{m}^{T}\widehat{\boldsymbol{\alpha}}_{\beta}-d).
\label{W2}%
\end{equation}
\end{definition}
\vspace{3pt}

In the following theorem, the asymptotic distribution of $Z_{K}(\widehat{\boldsymbol{\alpha}}_{\beta})$ is presented.

\vspace{3pt}
\begin{theorem}
The asymptotic distribution of  $Z$-type test statistics, $Z_{K} (\widehat{\boldsymbol{\alpha}}_{\beta})$, defined in (\ref{W2}), is standard normal.
\end{theorem}
\vspace{3pt}

Based on the previous result, the null hypothesis given in (\ref{W1}) will be rejected, with significance level $\alpha$,\ if $\left\vert Z_{K}(\widehat{\boldsymbol{\alpha}}_{\beta})\right\vert >z_{\frac{\alpha}{2}}$, where $z_{\frac{\alpha}{2}}$ is a right hand side quantile of order $\frac{\alpha}{2}$ of a normal distribution. Now we are going to present a result in order to provide an approximation for the test statistic defined in
(\ref{W2}).

\vspace{3pt}
\begin{theorem}
Let $\boldsymbol{\alpha}^{\ast}\in\Theta$ be the true value of the parameter
$\boldsymbol{\alpha}$ so that
\[
\widehat{\boldsymbol{\alpha}}_{\beta}\overset{\mathcal{P}
}{\underset{K\mathcal{\rightarrow}\infty}{\longrightarrow}}\boldsymbol{\alpha
}^{\ast}\in\Theta,
\]
and $\boldsymbol{m}^{T}\boldsymbol{\alpha}^{\ast}\neq d$. Then, the approximated power function of the test statistic in (\ref{W2}) at $\boldsymbol{\alpha}^{\ast}$ is given by equation (\ref{W31}), where $\Phi(\cdot)$ is the standard normal distribution function.

\begin{equation}
\pi\left(  \boldsymbol{\alpha}^{\ast}\right)  \simeq2\left(  1-\Phi\left(
z_{\frac{\alpha}{2}}-\sqrt{\frac{K}{\boldsymbol{m}^{T}\boldsymbol{\bar{J}%
}_{\beta}^{-1}(\boldsymbol{\alpha}^{\ast})\boldsymbol{\bar{K}}_{\beta
}(\boldsymbol{\alpha}^{\ast})\boldsymbol{\bar{J}}_{\beta}^{-1}%
(\boldsymbol{\alpha}^{\ast})\boldsymbol{m}}}(\boldsymbol{m}^{T}%
\boldsymbol{\alpha}^{\ast}-d)\right)  \right)  \label{W31}%
\end{equation}
\end{theorem}

\begin{remark}
Based on the previous results, it is possible to establish an explicit
expression of the number of devices%
\begin{footnotesize}
\[
K=\left[  \frac{\boldsymbol{m}^{T}\boldsymbol{\bar{J}}_{\beta}^{-1}%
(\boldsymbol{\alpha}^{\ast})\boldsymbol{\bar{K}}_{\beta}(\boldsymbol{\alpha
}^{\ast})\boldsymbol{\bar{J}}_{\beta}^{-1}(\boldsymbol{\alpha}^{\ast
})\boldsymbol{m}}{\boldsymbol{m}^{T}\boldsymbol{\alpha}^{\ast}-d}\left(
z_{\frac{\alpha}{2}}-\Phi^{-1}(1-\tfrac{\pi^{\ast}}{2})\right)  ^{2}\right]
+1,
\]
\end{footnotesize}

\noindent placed under temperatures $w_{i}$, $i=1,...,I$, at each time, $t_{j}$,
$j=1,2,...,J$, necessary in order to get a fixed power $\pi^{\ast}$ for a
specific significance level $\alpha$. Here, $[m]$ denotes  $\left[  \cdot\right]  $ the
largest integer less than or equal to $m$.
\end{remark}
\vspace{3pt}

\section{Real data examples\label{sec5}}
In this section, we present some numerical examples to illustrate the inferential results developed in the precedings sections. The first one is  an application to the reliability example considered in Section  \ref{sec2}, and the other two are real applications to tumorigenicity experiments considered earlier by other authors.

\subsection{Example 1 (Reliability experiment)}
Based on the example introduced in Section \ref{sec2}, in this section, the
MDPDEs of the parameters of the one-shot device model are
considered. As tuning parameter, $\beta\in\{0,$ $0.1,$ $0.2,$ $0.3,$ $0.4,$
$0.5,$ $0.6,$ $0.7,$ $0.8,$ $0.9,$ $1,\;2,\;3,\;4\}$ are taken. In Table
\ref{table2}, apart from the MDPDEs of $\boldsymbol{\alpha}$, the MDPDEs of
the reliability function%
\[
R(t|\lambda_{w_{0}}(\boldsymbol{\alpha}))=1-F(t|\lambda_{w_{0}}%
(\boldsymbol{\alpha}))=e^{-\lambda_{w_{0}}t}=\exp(-\alpha_{0}e^{\alpha
_{1}w_{0}}t)
\]
are also presented at  mission times (time points in the future at which we are interested
in the reliability of the unit) $t\in\{10,20,30\}$, namely $R(10|\lambda_{w_{0}}%
(\widehat{\boldsymbol{\alpha}}_{\beta}))$, $R(20|\lambda_{w_{0}}%
(\widehat{\boldsymbol{\alpha}}_{\beta}))$, $R(30|\lambda_{w_{0}}%
(\widehat{\boldsymbol{\alpha}}_{\beta}))$, as well as the MDPDEs of the mean
of the lifetime $T(\lambda_{w_{0}}(\boldsymbol{\alpha}))$, namely,
\[
E[T(\lambda_{w_{0}}(\boldsymbol{\alpha}))]=\frac{1}{\lambda_{w_{0}%
}(\boldsymbol{\alpha})}=\frac{1}{\alpha_{0}e^{\alpha_{1}w_{0}}},
\]
 under the normal operating temperature
$w_{0}=25$.%

 Table \ref{table2} shows that the mean lifetime obtained by the maximum likelihood estimator ($\beta=0$) is greater than that obtained from the alternative MPDPDEs. 

\begin{table}[h!!]   \centering
\renewcommand{\arraystretch}{1.3}
\caption{MDPDEs of the parameters, the reliability function at times $t\in\{10,$
$20,$ $30\}$, and  mean of lifetime at normal temperature of $25^{\circ}C$ in one-shot device testing experiment considered by Balakrishnan and Ling (2012). \label{table2}}%
\begin{tabular}
[c]{l||cc|cccc}%
$\beta$ & $\qquad\widehat{\alpha}_{0,\beta}\qquad$ & $\qquad\widehat{\alpha
}_{1,\beta}\qquad$ & $R(10|\lambda_{25}(\widehat{\boldsymbol{\alpha}}_{\beta
}))$ & $R(20|\lambda_{25}(\widehat{\boldsymbol{\alpha}}_{\beta}))$ &
$R(30|\lambda_{25}(\widehat{\boldsymbol{\alpha}}_{\beta}))$ & $E[T(\lambda
_{25}(\widehat{\boldsymbol{\alpha}}_{\beta}))]$\\\hline\hline
0 & 0.00487 & 0.04732 & 0.85300 & 0.72761 & 0.62065 & 62.89490\\
0.1 & 0.00489 & 0.04722 & 0.85288 & 0.72741 & 0.62039 & 62.83953\\
0.2 & 0.00490 & 0.04714 & 0.85277 & 0.72722 & 0.62016 & 62.79031\\
0.3 & 0.00491 & 0.04706 & 0.85268 & 0.72706 & 0.61995 & 62.74654\\
0.4 & 0.00492 & 0.04700 & 0.85260 & 0.72693 & 0.61978 & 62.70965\\
0.5 & 0.00493 & 0.04695 & 0.85253 & 0.72681 & 0.61963 & 62.67944\\
0.6 & 0.00494 & 0.04690 & 0.85247 & 0.72671 & 0.61950 & 62.65188\\
0.7 & 0.00495 & 0.04687 & 0.85246 & 0.72669 & 0.61947 & 62.64457\\
0.8 & 0.00495 & 0.04683 & 0.85236 & 0.72651 & 0.61925 & 62.59732\\
0.9 & 0.00496 & 0.04681 & 0.85233 & 0.72646 & 0.61918 & 62.58398\\
1 & 0.00496 & 0.04681 & 0.85239 & 0.72656 & 0.61931 & 62.61131\\
2 & 0.00496 & 0.04679 & 0.85231 & 0.72644 & 0.61915 & 62.57739\\
3 & 0.00494 & 0.04687 & 0.85255 & 0.72684 & 0.61966 & 62.68584\\
4 & 0.00491 & 0.04700 & 0.85292 & 0.72748 & 0.62048 & 62.85869\\\hline
\end{tabular}
$\ \ \ \ \ $%
\end{table}%
\subsection{Example 2 (ED01 Data)}

In 1974,  the National Center for Toxicological Research made an experiment on 24000 female mice  randomized to a control group  or one of seven dose levels of a known carcinogen, called  2-Acetylaminofluorene (2-AAF). Table 1 in Lyndsey and Ryan (1993) shows the results obtained when the highest dose level ($150$ parts per million) was administered.  The original study considered four different outcomes: Number of animals dying tumour free  (DNT) and with tumour (DWT), and sacrified without tumour (SNT) and with tumour (SWT), summarized over three time intervals at $12$, $18$ and $33$ months. A total of $3355$ mice were involved in the experiment.
\begin{table}[!h]
\renewcommand{\arraystretch}{1.3}
\centering
\caption{Number of mice  sacrified  ($r=0$) and died without tumour ($r=1$) and with tumour ($r=2$)    from the ED01 Data} \label{table:ED01}
\begin{tabular}{ llccc }
\multicolumn{5}{ c }{} \\
\hline
 & & $r=0$ & $r=1$& $r=2$ \\ \hline
\multirow{2}{*}{$IT_1=12$} & $w=0$ & 115 & 22 &8\\
 & $w=1$ & 110 &  49& 16 \\ \hline
\multirow{2}{*}{$IT_2=18$} & $w=0$ & 780 & 42 & 8\\
 & $w=1$ & 540 & 54 & 26 \\ \hline

\multirow{2}{*}{$IT_3=33$} & $w=0$ & 675 & 200 & 85 \\
 & $w=1$ & 510 & 64 & 51 \\
\hline
\end{tabular}
\end{table}

Balakrishnan et al. (2016a) made an analysis combining  SNT and SWT as the sacrificed group ($r=0$);
and denoting the cause of DNT as natural death ($r=1$), and the cause of DWT as death due to cancer ($r=2$). This modified data are presented in Table \ref{table:ED01}, while MDPDEs of the model parameters and the corresponding  estimates of mean lifetimes are presented in Table \ref{table:MPDE_ED01}. Here $w=0$ refers to control group and $w=1$ is the test group, while $E(T_1)$ and $E(T_2)$  are the estimated mean lifetimes for sacrifice or nature death ($r=0,1$) and death due to cancer ($r=2$), respectively.

\begin{table}[!h]
\renewcommand{\arraystretch}{1.3}
\centering
\centering
\caption{ MDPDEs of the parameters and the mean lifetimes of the ED01 experiment}\label{table:MPDE_ED01}
\footnotesize

\vspace{0.4cm}
\begin{tabular}{|l|rrrr|rrrr|rr|}
 \hline
$\beta$ & $\widehat{\alpha}_{10}$ & $\widehat{\alpha}_{11}$ & E$_{w=0}$($T_1$) & E$_{w=1}$($T_1$) & $\widehat{\alpha}_{20}$ & $\widehat{\alpha}_{21}$ & E$_{w=0}$($T_2$) & E$_{w=1}$($T_2$) & E$_{w=0}$($T$) & E$_{w=1}$($T$)  \\ \hline
0    & 0.00617   & $-$0.12790   & 162.233  & 184.165   & 0.00236   & 0.25620    & 426.425  & 331.582  & 117.447 & 118.299 \\
0.1  & 0.00702   & 0.09355   & 142.352  & 129.639   & 0.00250    & 0.32870    & 399.794  & 287.795 & 104.988 & 89.392  \\
0.2  & 0.00698   & 0.06495   & 143.302  & 134.290  & 0.00250    & 0.31173   & 400.433 & 293.189  & 105.504 & 92.072  \\
0.3  & 0.00703   & 0.00999   & 142.253  & 140.840    & 0.00249   & 0.29613   & 401.393   & 298.513 & 105.045 & 95.708  \\
0.4  & 0.00690    & 0.00998   & 145.019  & 143.578  & 0.00249   & 0.27957   & 401.602  & 303.655   & 106.545 & 97.484  \\
0.5  & 0.00677   & 0.00998   & 147.662  & 146.195  & 0.00249   & 0.26421   & 401.839   & 308.537  & 107.965 & 99.175  \\
0.6  & 0.00666   & 0.00998   & 150.085  & 148.594  & 0.00283   & 0.00997   & 353.925  & 350.414  & 105.342 & 104.296 \\
0.7  & 0.00682   & $-$0.06678  & 146.635 & 156.763  & 0.00249   & 0.23702   & 401.985  & 317.157  & 107.415 & 104.876 \\
0.8  & 0.00680    & $-$0.08753  & 147.020  & 160.468  & 0.00279   & 0.00997   & 358.642   & 355.083 & 104.256 & 110.508 \\
0.9  & 0.00679   & $-$0.10530   & 147.321  & 163.680  & 0.00278   & 0.00997   & 360.357  & 356.781  & 104.516 & 112.141\\
1    & 0.00678   & $-$0.11980   & 147.546  & 166.324  & 0.00277   & 0.00995   & 361.607   & 358.028 & 104.739 & 113.506 \\  \hline
\end{tabular}
\end{table}

\begin{figure}[h!!!!]
\centering
\includegraphics[scale=0.7]{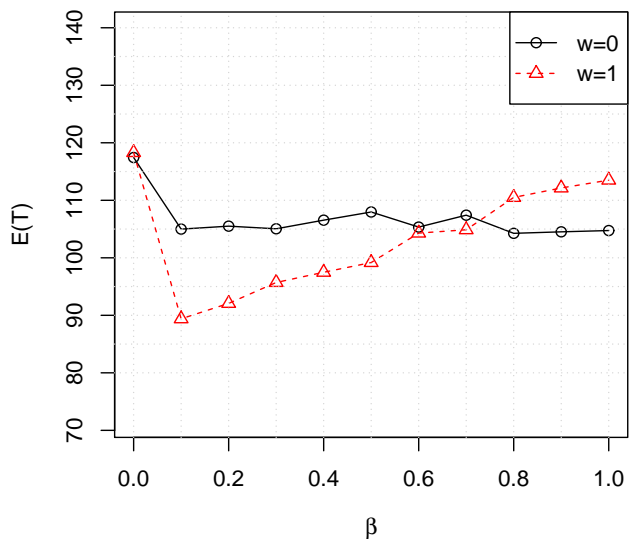}
\caption{ MDPDEs of the mean lifetimes, for different values of the tuning parameter $\beta$, from the ED01 experiment  } \label{fig: lifetimeT_ED01}
\end{figure}

From Table \ref{table:MPDE_ED01}, some MDPDEs of $\alpha_{11}$  are seen to be negative. As pointed out in Balakrishnan et al. (2016), this can be due to the fact that the true value of it may be quite close to zero. In fact, for the values of the tuning parameter $\beta \in \{0.1,0.2,0.3,0.4,0.5,0.6,0.7 \}$, the estimators of $\alpha_{11}$ are very close to zero, meaning that the drug will not increase the hazard rate of the natural death outcome. Furthermore,  if we look at the estimates of mean lifetimes, these last estimators show a reduction when the carcinogenic drug is administered, but the other ones, $\beta \in \{0,0.8,0.9,1 \}$, do not show this behavior (see Figure \ref{fig: lifetimeT_ED01}). Thus, in this case, we observe that the MDPDEs with tuning parameter $\beta \in \{0.1,0.2,0.3,0.4,0.5,0.6,0.7 \}$  give a more meaningful result in the context of the laboratory experiment than, in particular,  the maximum likelihood estimator ($\beta=0$). The simulation study presented in this paper will prove how, in a general case,  MDPDEs with these tuning parameters will also present a better behaviour in terms of robustness.

\subsection{Example 3 (Benzidine Dihydrochloride Data)}
\begin{table}[h!!!!!]

\centering
\caption{Number of mice  sacrified  ($r=0$) and died without tumour ($r=1$) and with tumour ($r=2$)    from the Benzidine Dihydrochloride Data} \label{table:Benzidine}
\begin{tabular}{ llccc }
\multicolumn{5}{ c }{} \\
\hline
 & & $r=0$ & $r=1$& $r=2$ \\ \hline
\multirow{2}{*}{$IT_1=9.37$} & $w=1$ & 70   & 2 &0\\
 & $w=2$ & 22 & 3 & 0 \\ \hline
\multirow{2}{*}{$IT_2=14.07$} & $w=1$ &48 & 1 & 0\\
 & $w=2$ & 14 & 4 & 17 \\ \hline

\multirow{2}{*}{$IT_3=18.7$} & $w=1$ & 35 & 4 & 7 \\
 & $w=2$ & 1 & 1 & 9 \\
\hline
\end{tabular}
\end{table}

The benzidine dihydrochloride experiment was also conducted  at the National Center for Toxicological Research to examine  the incidence in mice of liver tumours  induced by the drug, and studied  by Lyndsey and Ryan (1993) and   Balakrishnan et al. (2016b).  The inspection times used on the mice were $9.37$, $14.07$ and $18.7$ months. In Table \ref{table:MPDE_Benzidine}, the numbers of mice sacrified ($r=0$), died without tumour ($r=1$) and died with tumour ($r=2$), are shown, for two different doses of drug: $60$ parts per million ($w=1$) and $400$ parts per million ($w=2$). As in the previous example, we consider as ``failures'' the  mice died due to cancer.

Table \ref{table:MPDE_Benzidine} shows the MDPDEs of the model parameters and the corresponding  estimates of mean lifetimes. Although some differences are observed in the results for different values of the tuning parameter, in all the cases, the mean lifetime shows a reduction when the carcinogenic drug is administered.

\begin{table}[h!]
\renewcommand{\arraystretch}{1.3}
\centering
\caption{ MDPDEs of the parameters and the mean lifetimes of the Benzidine Dihydrochlorid experiment}\label{table:MPDE_Benzidine}
\footnotesize
\vspace{0.4cm}
\begin{tabular}{|l|rrrr|rrrr|rr|}
 \hline
$\beta$ & $\widehat{\alpha}_{10}$ & $\widehat{\alpha}_{11}$ & E$_{w=0}$($T_1$) & E$_{w=1}$($T_1$) & $\widehat{\alpha}_{20}$ & $\widehat{\alpha}_{21}$ & E$_{w=1}$($T_2$) & E$_{w=2}$($T_2$) & E$_{w=1}$($T$) & E$_{w=2}$($T$)  \\ \hline
0         & 0.00074   & 1.08665   & 1342.580 & 452.912  & 0.00018   & 2.49999   & 5472.201  & 449.190 & 1081.274 & 227.233 \\
0.1       & 0.00093   & 0.87121   & 1072.790 & 448.905  & 0.00022   & 2.45781   & 4459.410   & 381.825 & 867.943  & 208.460 \\
0.2       & 0.00097   & 0.84038   & 1032.863 & 445.729  & 0.00024   & 2.42125   & 4110.686  & 365.071  & 827.690    & 202.187 \\
0.3       & 0.00101   & 0.81098   & 994.958  & 442.182  & 0.00026   & 2.39084   & 3867.836  & 354.112 & 790.471  & 196.024 \\
0.4       & 0.00104   & 0.78168   & 958.766  & 438.766  & 0.00029   & 2.34614   & 3507.841  & 335.834 & 750.183  & 188.387 \\
0.5       & 0.00109   & 0.75071   & 920.459 & 434.483  & 0.00029   & 2.33901   & 3449.648  & 332.624 & 726.525  & 188.353 \\
0.6       & 0.00112   & 0.72656   & 893.899  & 432.261  & 0.00032   & 2.29717   & 3168.017  & 318.521 & 695.074  & 181.946 \\
0.7       & 0.00115   & 0.70252   & 866.492 & 429.206  & 0.00032   & 2.28271   & 3078.308  & 314.009 & 678.390  & 182.918 \\
0.8       & 0.00118   & 0.68232   & 845.366   & 427.285  & 0.00033   & 2.27346   & 3011.326  & 310.030 & 660.973  & 180.322 \\
0.9       & 0.00121   & 0.66476   & 826.449  & 425.122   & 0.00034   & 2.25372   & 2902.649  & 304.799 & 645.163  & 178.887 \\
1         & 0.00124   & 0.64796   & 807.541  & 422.432  & 0.00035   & 2.23942   & 2823.643  & 300.774 & 629.593 & 176.897\\ \hline
\end{tabular}
\end{table}

In order to have an idea of the behavior of the different MDPDEs, in relation to the efficiency as well as the robustness, we carry out an extensive simulation study in the next section.

\section{Simulation study\label{sec6}}

In this section, a simulation study is carried out to examine the behavior of the
MDPDEs of the parameters of the one-shot device model, studied in Section
\ref{sec3}, as well as the $Z$-type tests, based on MDPDEs, detailed  in
Section \ref{sec4}. We pay special attention to the robustness issue. It is
interesting to note, in this context, the following. For each fixed time,
$t_{j}$, under a fixed temperature, $w_{i}$, $K$ devices are tested. In this
sense, we can identify our data as a $I\times J$ contingency table with $K$
observations in each cell. Hence, under  this setting, we must consider
\textquotedblleft outlying cells\textquotedblright\ rather than
\textquotedblleft outlying observations\textquotedblright. A cell which does
not follow the one-shot device model will be called an outlying cell or
outlier. The strong outliers may lead to reject a model fitting even if the rest
of the cells fit the model properly. In other cases, even though the cells
seem to fit reasonably well the model, the outlying cells contribute to an increase
in the values of the residuals as well as the divergence measure between the data
and the fitted values according to the one-shot device model considered.
Therefore, it is very important to have robust estimators as well as robust
test statistics in order to avoid the undesirable effects of the outliers
in the data. The main purpose of this simulation study is to show
that inside the family of MDPDEs, developed here, there are
estimators with better robust properties than the MLE, and the $Z$-type tests
constructed from them are at the same time more robust than the classical
$Z$-type test,  constructed through the MLEs.

\subsection{The MDPDEs\label{sec6.1}}

In this section, we carry out a simulation study to compare the behavior of some MDPDEs with respect to the MLEs of the parameters in the one-shot device model under the exponential distribution. In order to evaluate th performance of the proposed MDPDEs, as well as the MLEs, we consider the root of the mean square errors (RMSEs). We have considered a model in which, $I=J=3$, $w\in\{35,$ $45,$ $55\}$, $t\in\{10,$ $20,$ $30\}$ and $K=20$, as in the example in Table \ref{table1}, and the simulation experiment proposed by Ling (2012). This model has been examined under three choices of $(\alpha_{0},\alpha_{1})=(0.005,0.05)$, $(\alpha_{0},\alpha_{1})=(0.004,0.05)$ and $(\alpha_{0},\alpha_{1})=(0.003,0.05)$ for low-moderate, moderate and moderate-high reliability, respectively.

\begin{figure}[!h!!!!!] 
\begin{subfigure} {0.5\textwidth}
\centering
\begin{tabular}[c]{c}
\raisebox{-0cm}{\includegraphics[scale=0.55]
{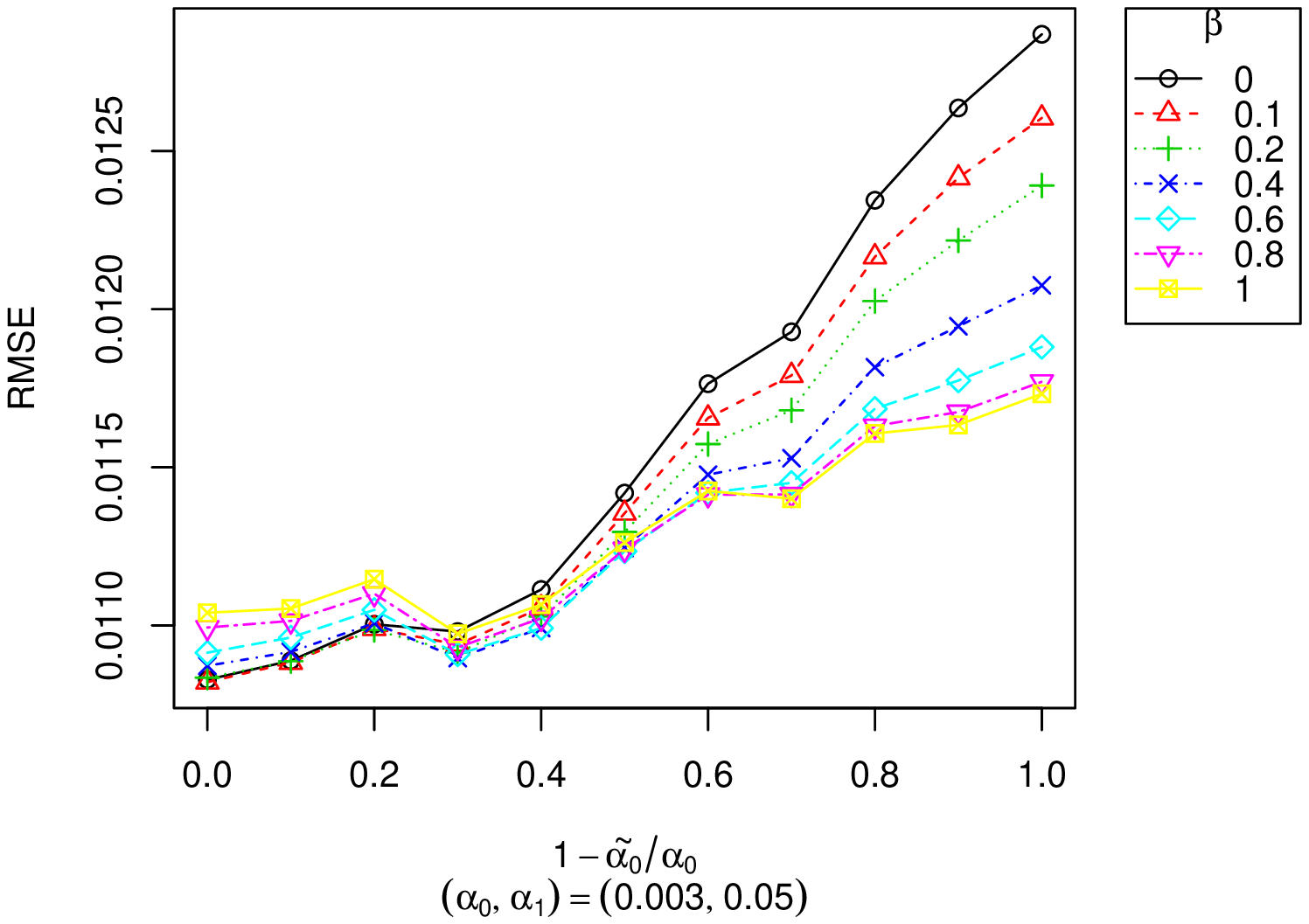}}
\\
\raisebox{-0cm}{\includegraphics[scale=0.55]
{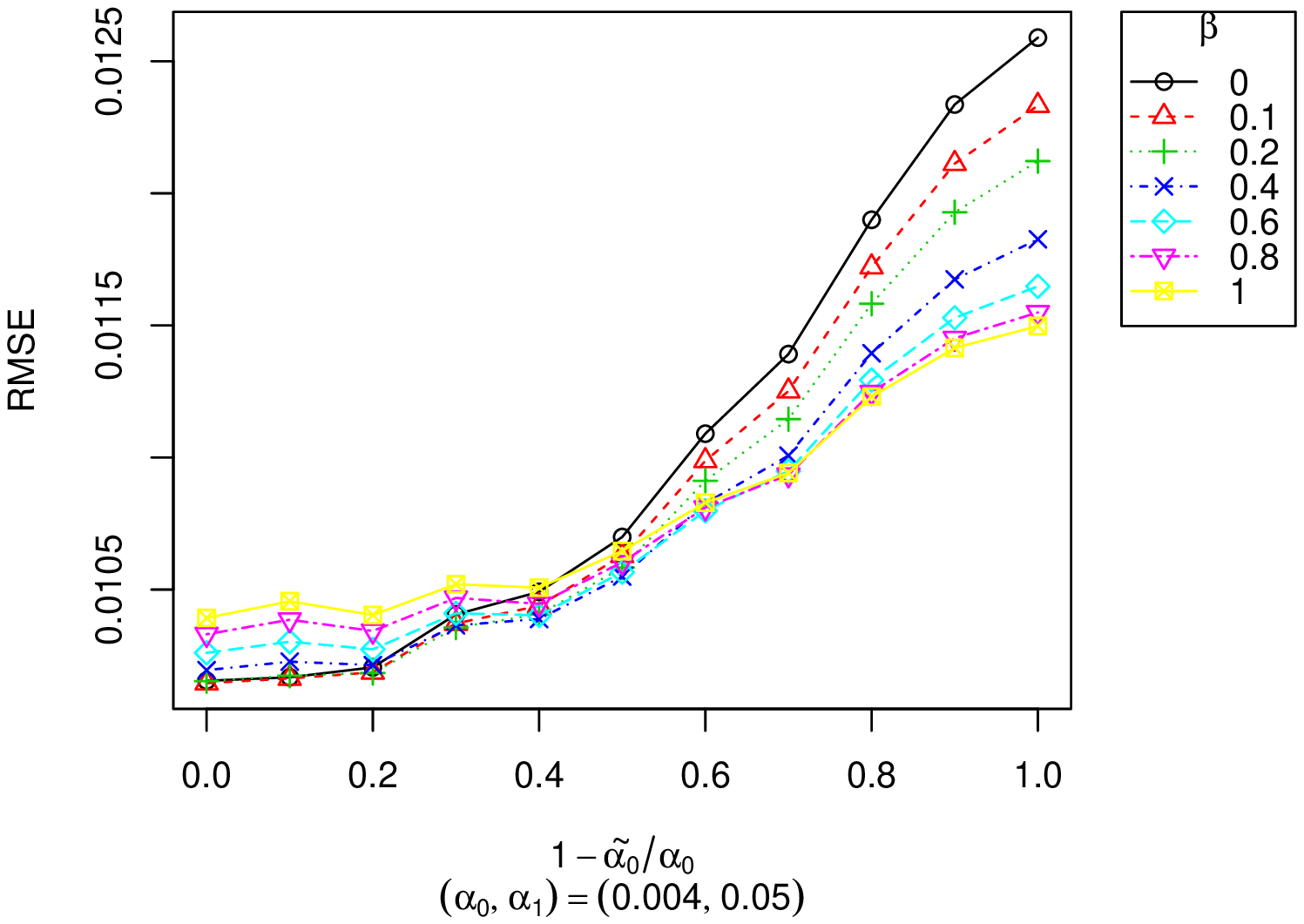}}
\\
\raisebox{-0cm}{\includegraphics[scale=0.55]
{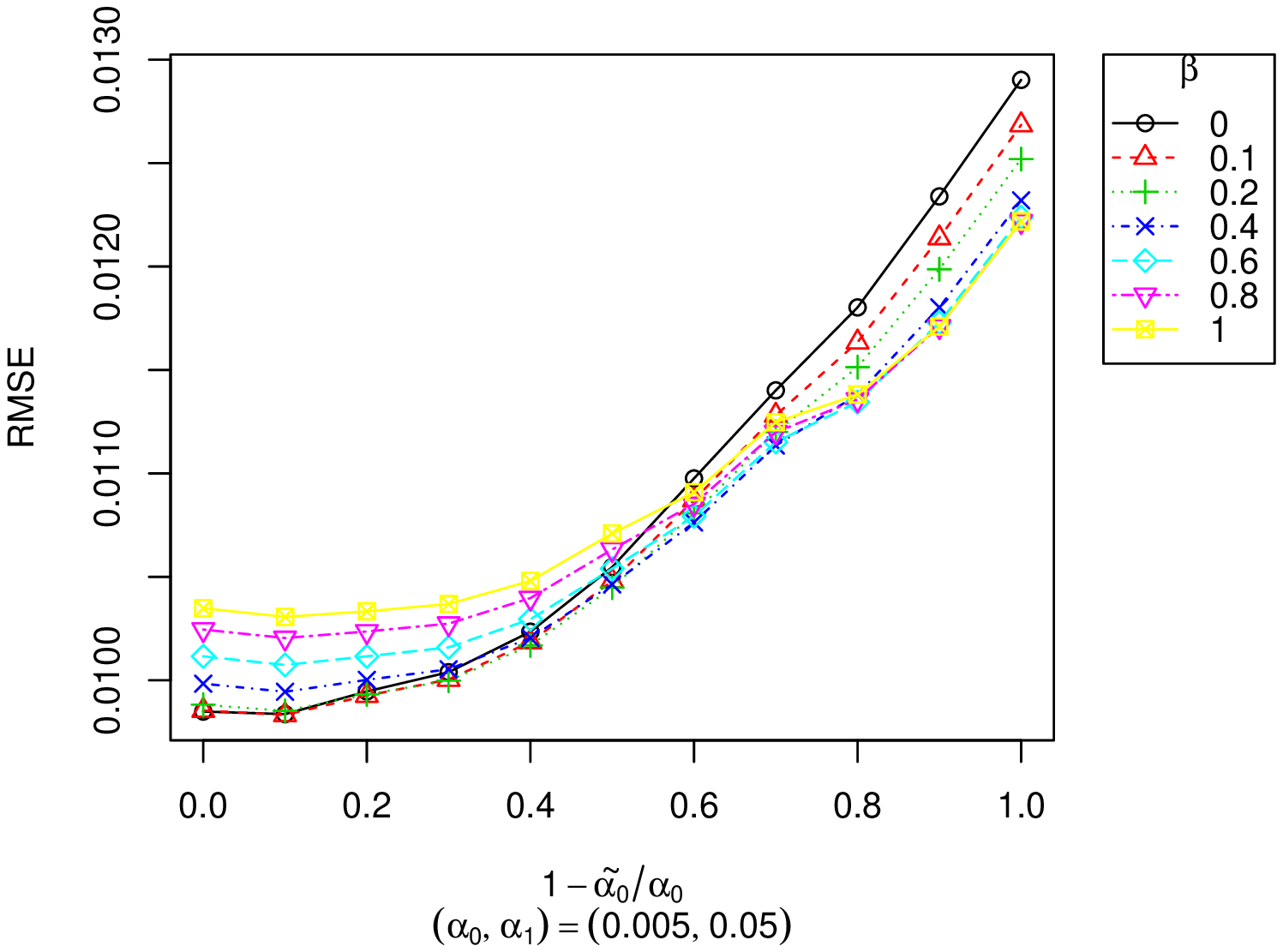}}
\end{tabular}
\caption{ with an $\alpha_0$-contaminated outlying cell.\label{fig:MPDE_1}}
\vspace*{1.5cm}

\end{subfigure}
\begin{subfigure} {0.5\textwidth}
\begin{tabular}
[c]{c}%
\raisebox{-0cm}{\includegraphics[scale=0.55]
{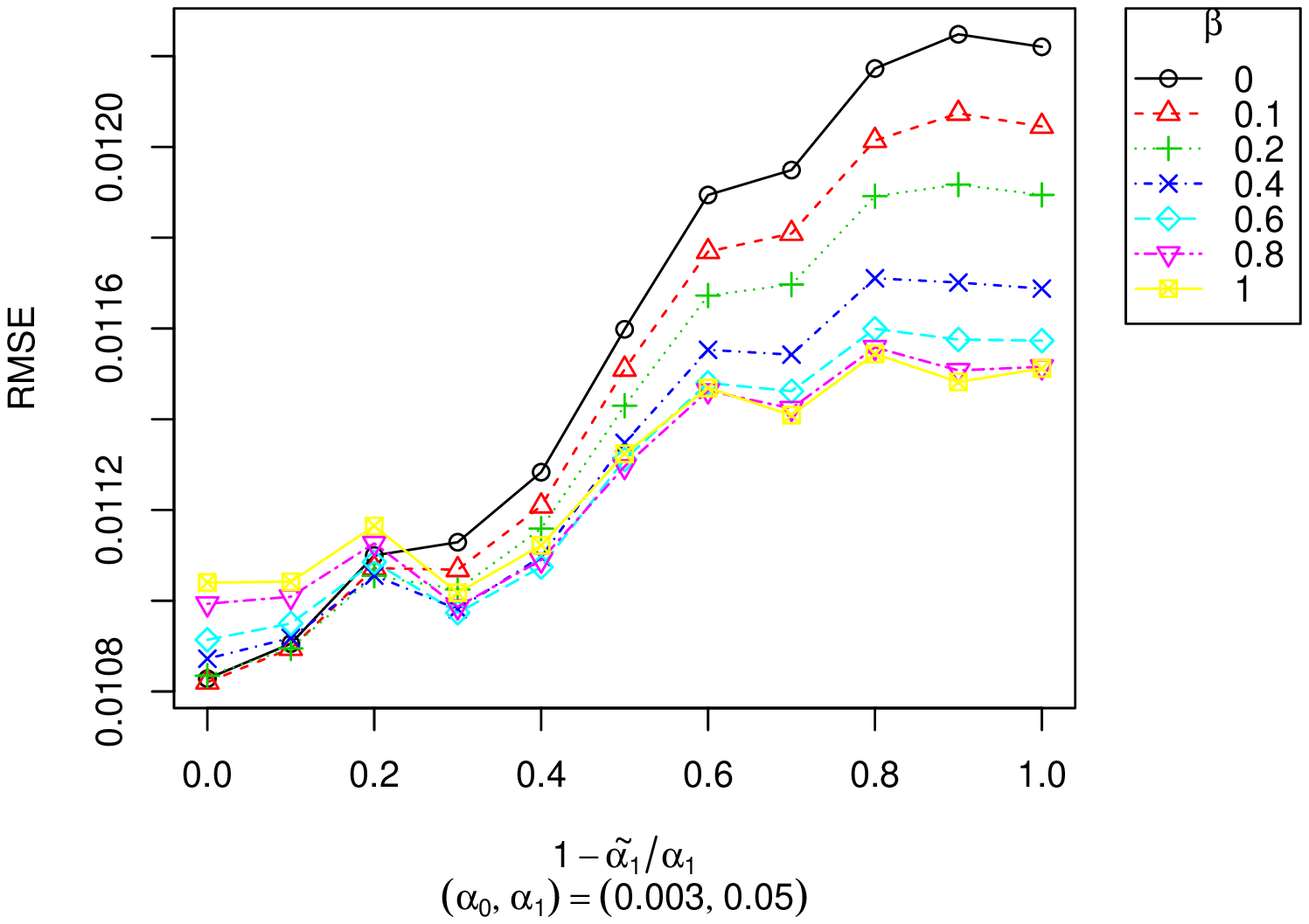}}
\\
\raisebox{-0cm}{\includegraphics[scale=0.55]
{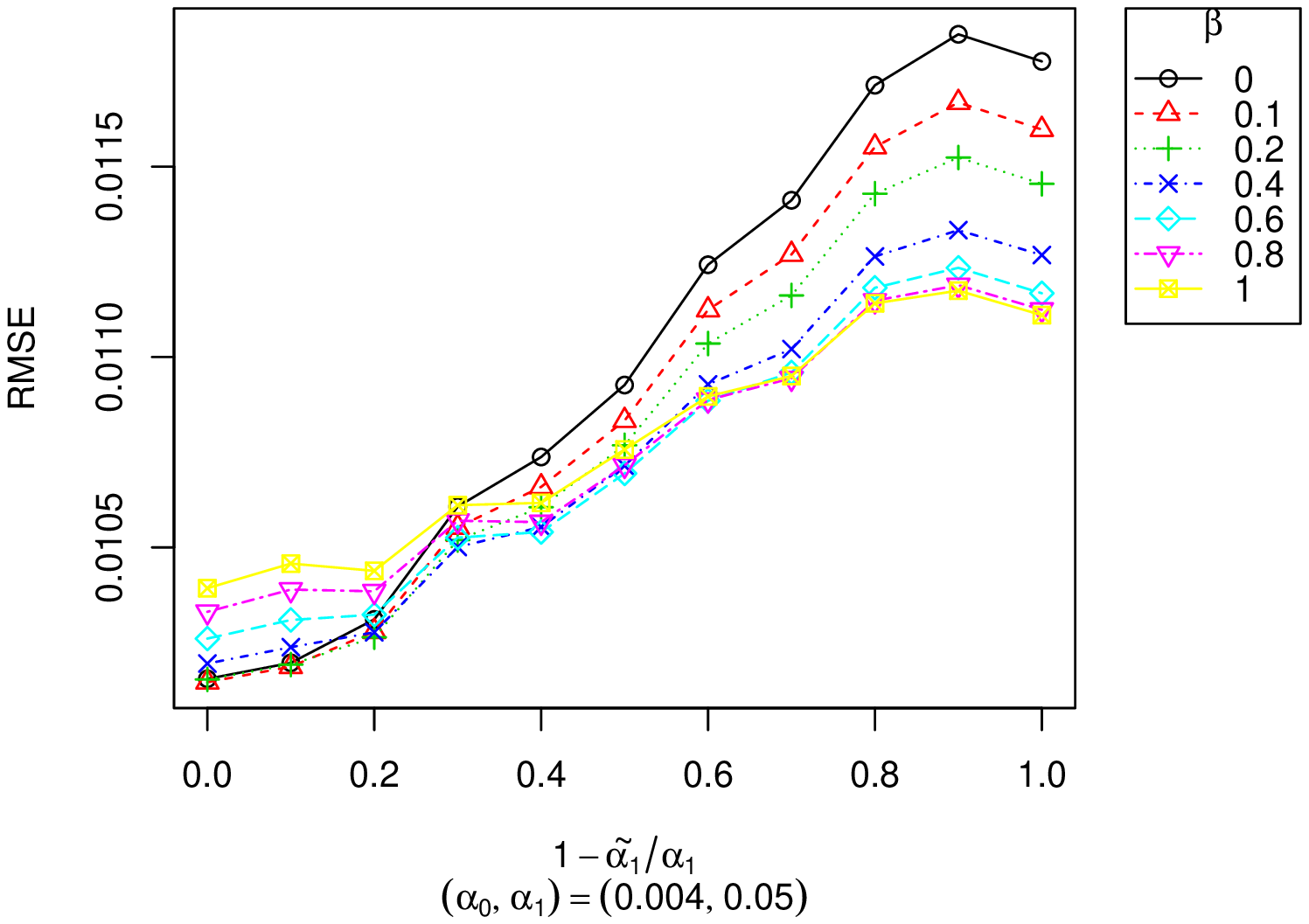}}
\\
\raisebox{-0cm}{\includegraphics[scale=0.55]
{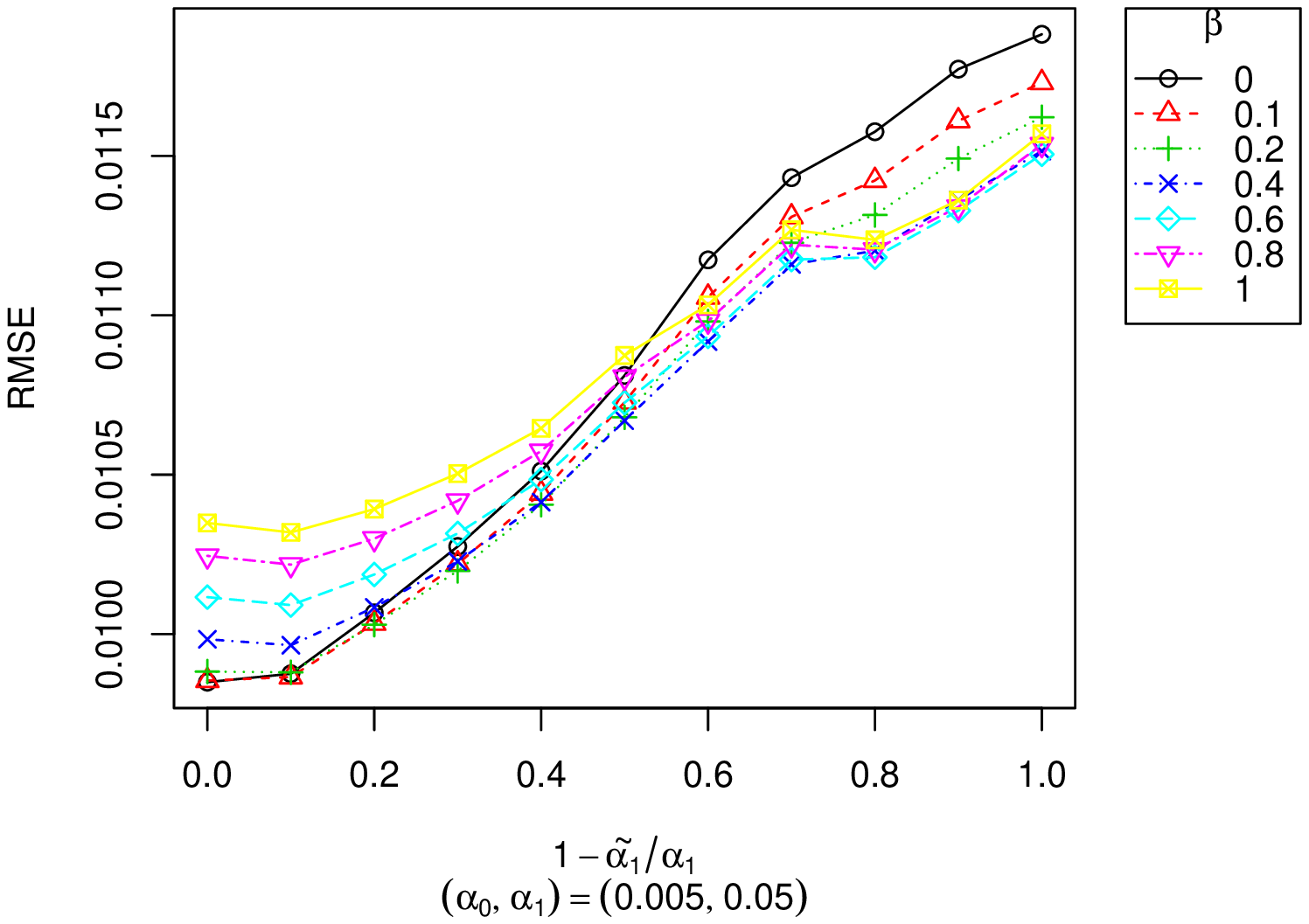}}
\end{tabular}
\caption{ with an $\alpha_1$-contaminated outlying cell.\label{fig:MPDE_2}}
\vspace*{1.5cm}

\end{subfigure}%
\caption{RMSEs of MDPDEs for $\boldsymbol{\alpha }$}
\end{figure}

To evaluate the robustness of the MDPDEs, we have studied the behavior of this model under the consideration of an outlying cell for $(w_{1},t_{1})$ in our contingency table, with $10,000$ replications and estimators corresponding to the tuning parameter $\beta\in\{0,0.1,0.2,0.4,0.6,0.8,1\}$. The reduction of each one of the parameters of the outlying cell, denoted by $\tilde{\alpha}_{0}$ or $\tilde{\alpha}_{1}$ ($\alpha_{0}\geq\tilde{\alpha}_{0}$ or $\alpha_{1}\geq\tilde{\alpha}_{1}$) increases the mean of its lifetime distribution function in (\ref{eq:expo}). The results obtained by decreasing parameter $\alpha_{0}$ are shown in Figure \ref{fig:MPDE_1}, while the results obtained by decreasing parameter $\alpha_{1}$ are shown in Figure \ref{fig:MPDE_2}. In all the cases, we can see how the MLEs and the MDPDEs with small values of tuning parameter $\beta$ present the smallest RMSEs for weak outliers, i.e., when $\tilde{\alpha}_{0}$ is close to $\alpha_{0}$ ($1-\tilde{\alpha}_{0}/\alpha_{0}$ is close to $0$) or $\tilde{\alpha}_{1}$ is close to $\alpha_{1}$ ($1-\tilde{\alpha}_{1}/\alpha_{1}$ is close to $0$). On the other hand, large values of tuning parameter $\beta$ turn the MDPDEs to present the smallest RMSEs, for medium and strong outliers, i.e., when $\tilde{\alpha}_{0}$ is not close to $\alpha_{0}$ ($1-\tilde{\alpha}_{0}/\alpha_{0}$ is not close to $0$) or $\tilde{\alpha}_{1}$ is not close to $\alpha_{1}$ ($1-\tilde{\alpha}_{1}/\alpha_{1}$ is not close to $0$). Therefore, the MLE of $(\alpha_{0},\alpha_{1})$ is very efficient when there are no outliers, but highly non-robust when there are outliers. On the other hand, the MDPDEs with moderate values of the tuning parameter $\beta$ exhibit a little loss of efficiency without outliers but at the same time a considerable improvement of robustness with outliers. Actually, these values of the tuning parameter $\beta$ are the most appropriate ones for the estimators of the parameters in the one-shot device model according to robustness theory: To improve in a considerable way the robustness of the estimators,  a small amount of efficiency needs to be compromised.

\subsection{The Z-type tests based on MDPDEs\label{sec6.2}}

\begin{figure}[!h]  
\centering
\begin{tabular}[c]{cc}%
\raisebox{-0cm}{\includegraphics[scale=0.44]
{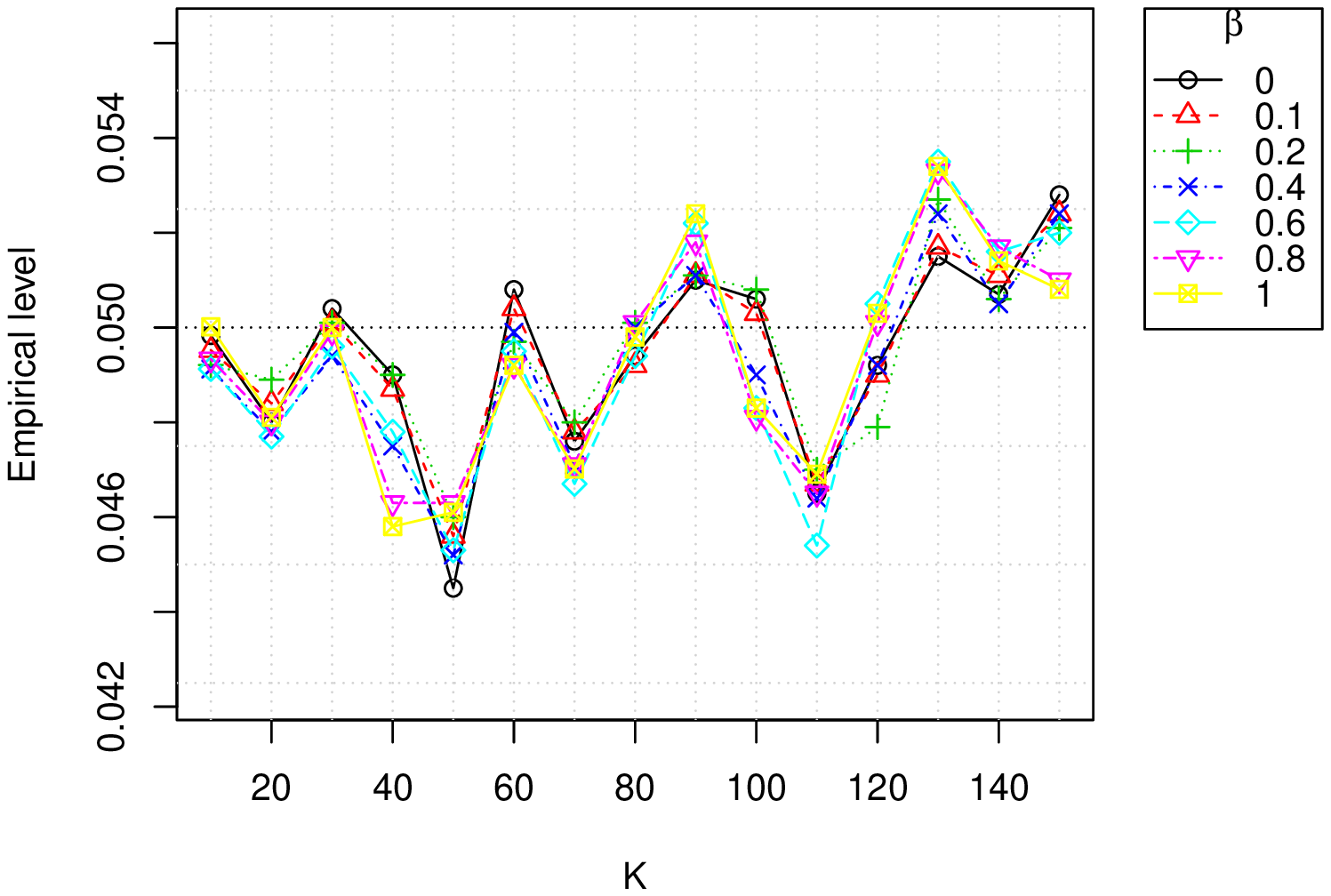}}%
&
\raisebox{-0cm}{\includegraphics[scale=0.44]
{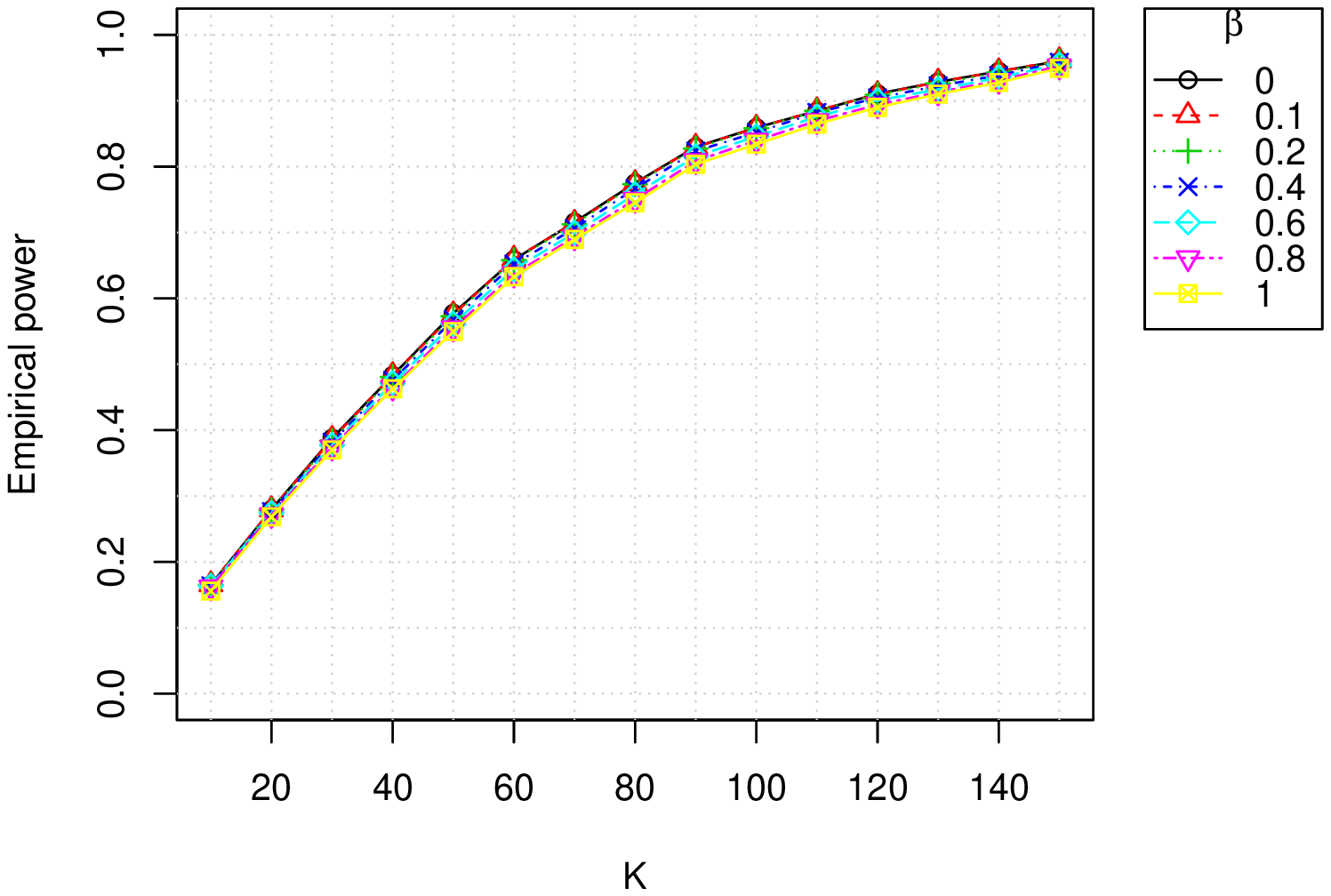}}
\end{tabular}
\caption{Simulated levels (left) and powers (right) with no outliers in the data.\label{fig:LP_1}}%
\end{figure}

We will study the performance, with respect to robustness, through simulation of the one-shot device model defined in Section \ref{sec2}\ with the same values of $I,J,t,w$ of the example of Balakrishnan and Ling (2012) given in Table \ref{table1}\ and for the same tuning parameter, $\beta$, as in Section \ref{sec6.1}. We are interested in testing the null hypothesis $H_{0}:\alpha_{1}=0.05$ against the alternative $H_{1}:\alpha_{1}\neq0.05$, through the $Z$-type test statistics based on MDPDEs. Under the null hypothesis, we consider as true parameters $(\alpha_{0},\alpha_{1})=(0.004,0.05)$, while under the alternative we consider as true parameters $(\alpha_{0},\alpha_{1})=(0.004,0.02)$. In Figure \ref{fig:LP_1}, we present the empirical significance level (measured as the proportions of test statistics exceeding in absolute value the standard normal quantile critical value) with $10,000$ replications. The empirical power (obtained in a similar manner) is also presented in the right hand side of Figure \ref{fig:LP_1}. Notice that in all the cases the observed levels are quite close to the nominal level of $0.05$. The empirical power is similar for the different values of the tuning  parameters $\beta$, a bit lower for large values of $\beta$, and closer to one as $K$ or the sample size ($n=IJK$) increases.

\begin{figure}[h] 
 \centering
\begin{tabular}
[c]{cc}
\raisebox{-0cm}{\includegraphics[scale=0.54]
{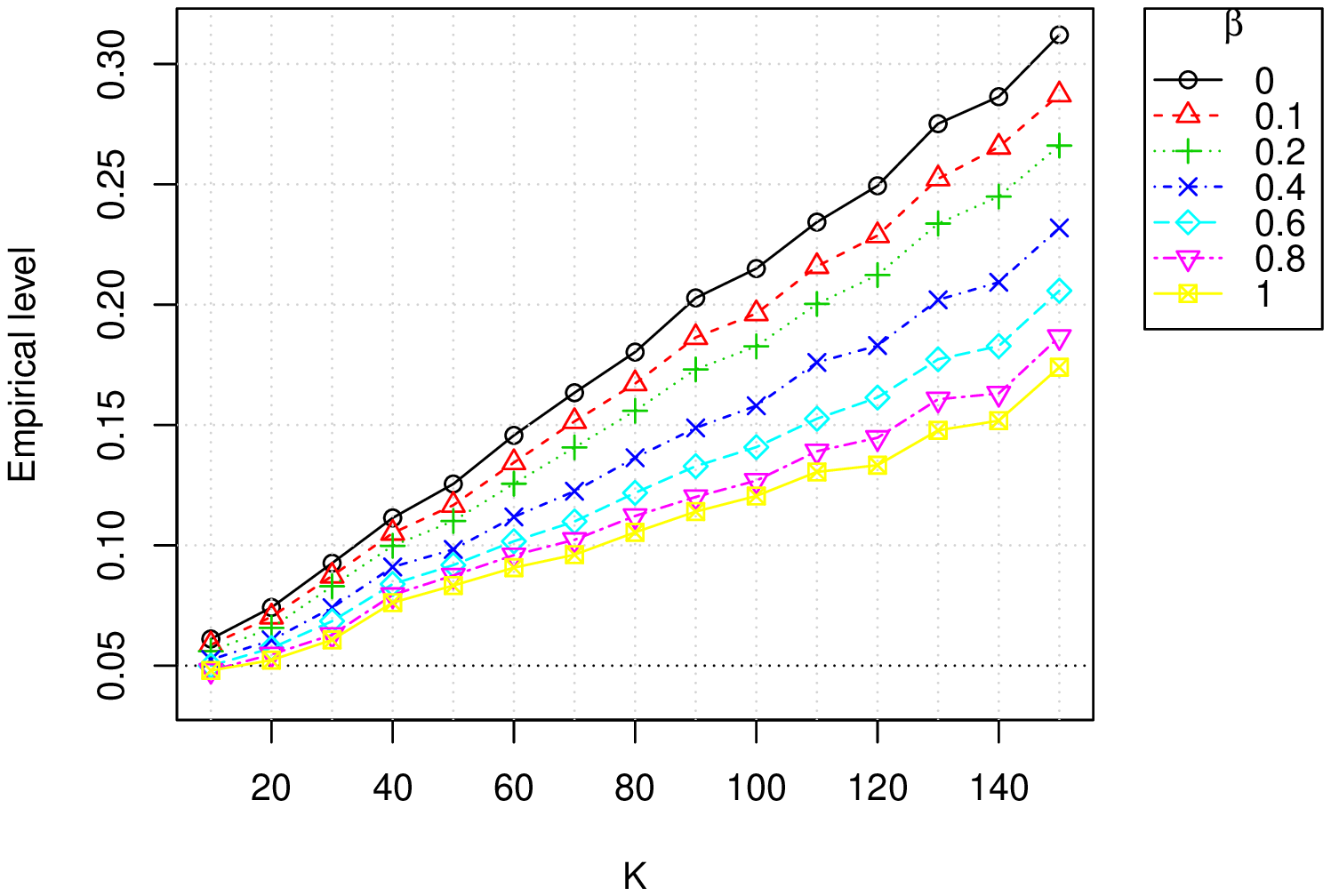}}
&
\raisebox{-0cm}{\includegraphics[scale=0.54]
{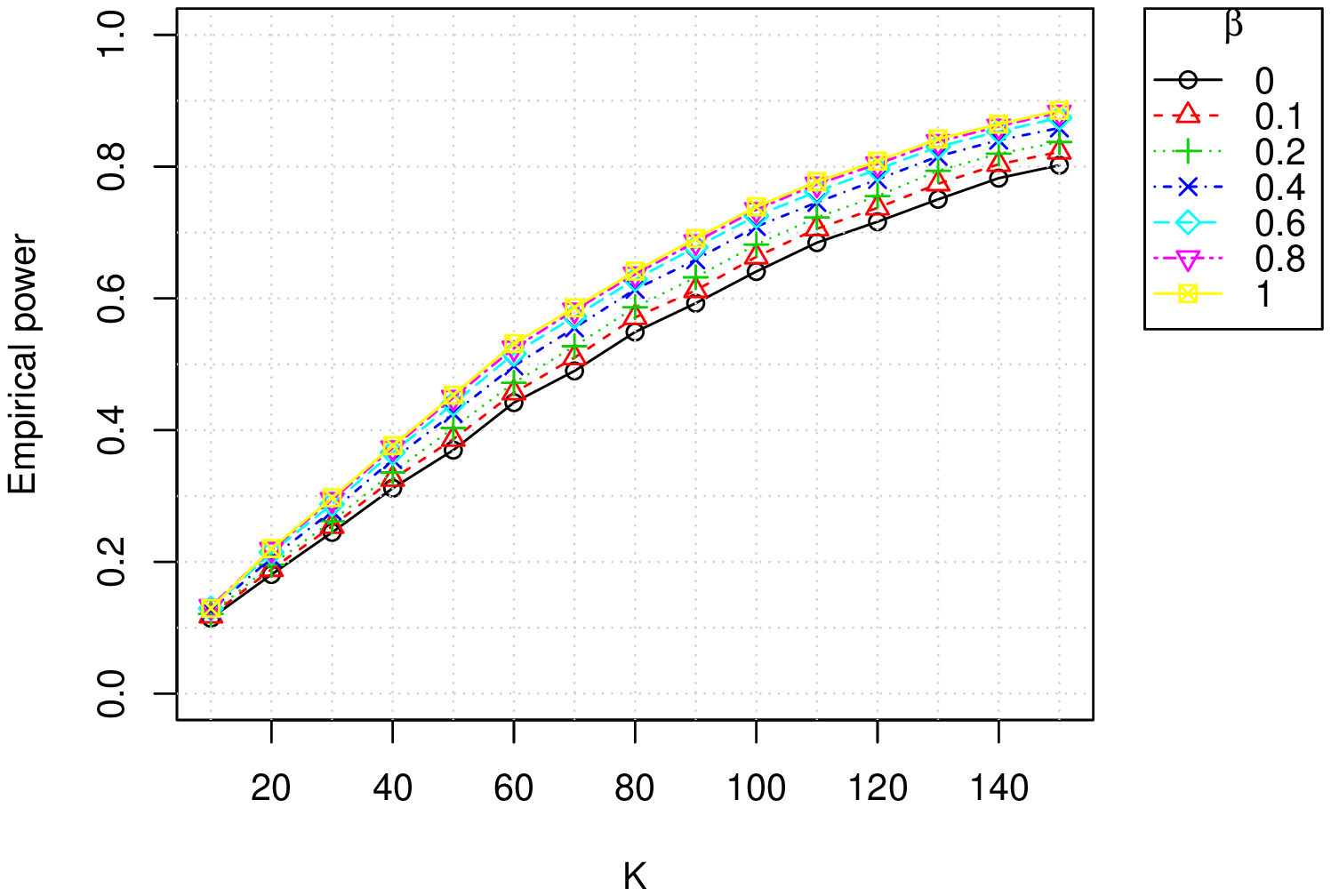}}
\\
\raisebox{-0cm}{\includegraphics[scale=0.54]
{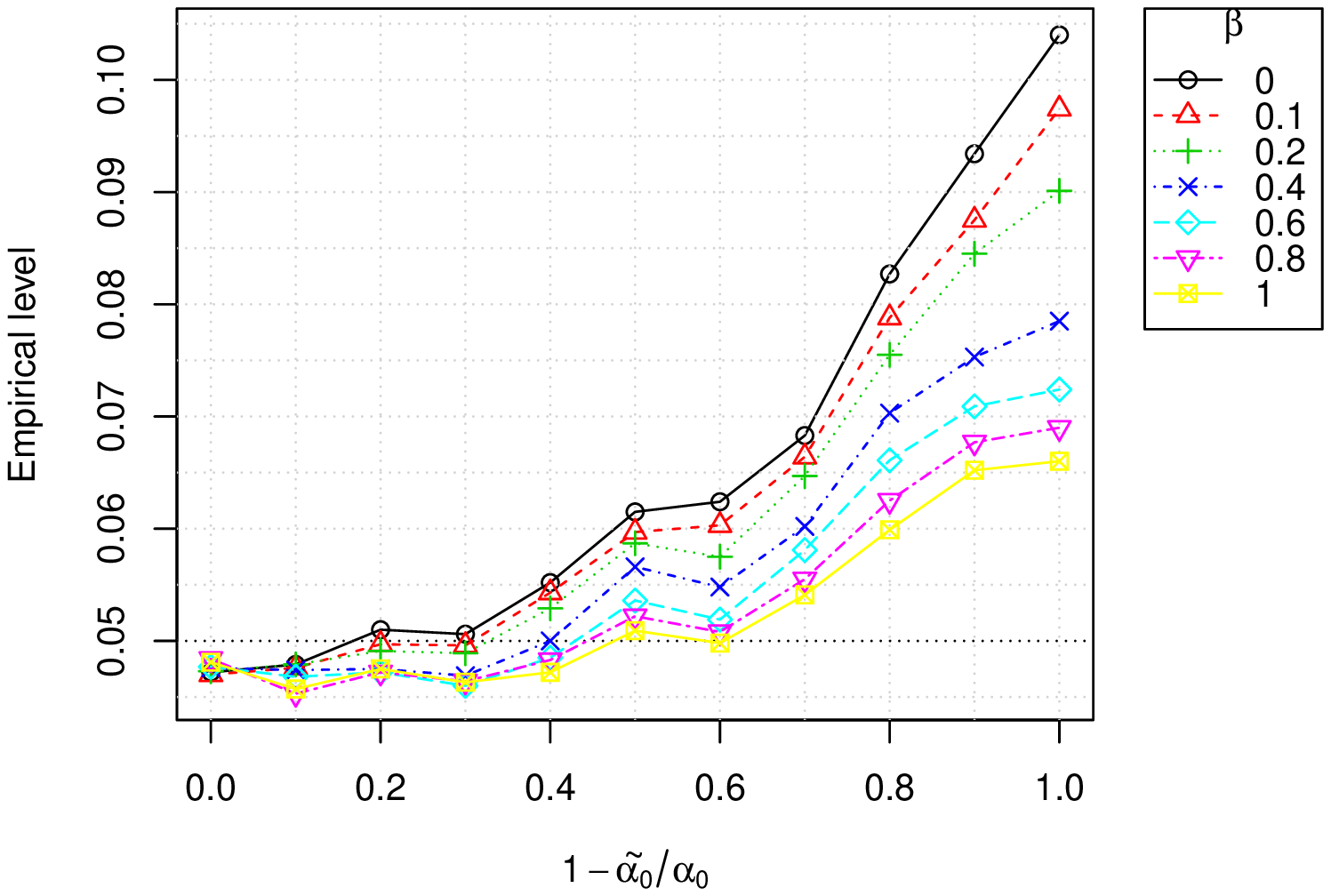}}
&
\raisebox{-0cm}{\includegraphics[scale=0.54]
{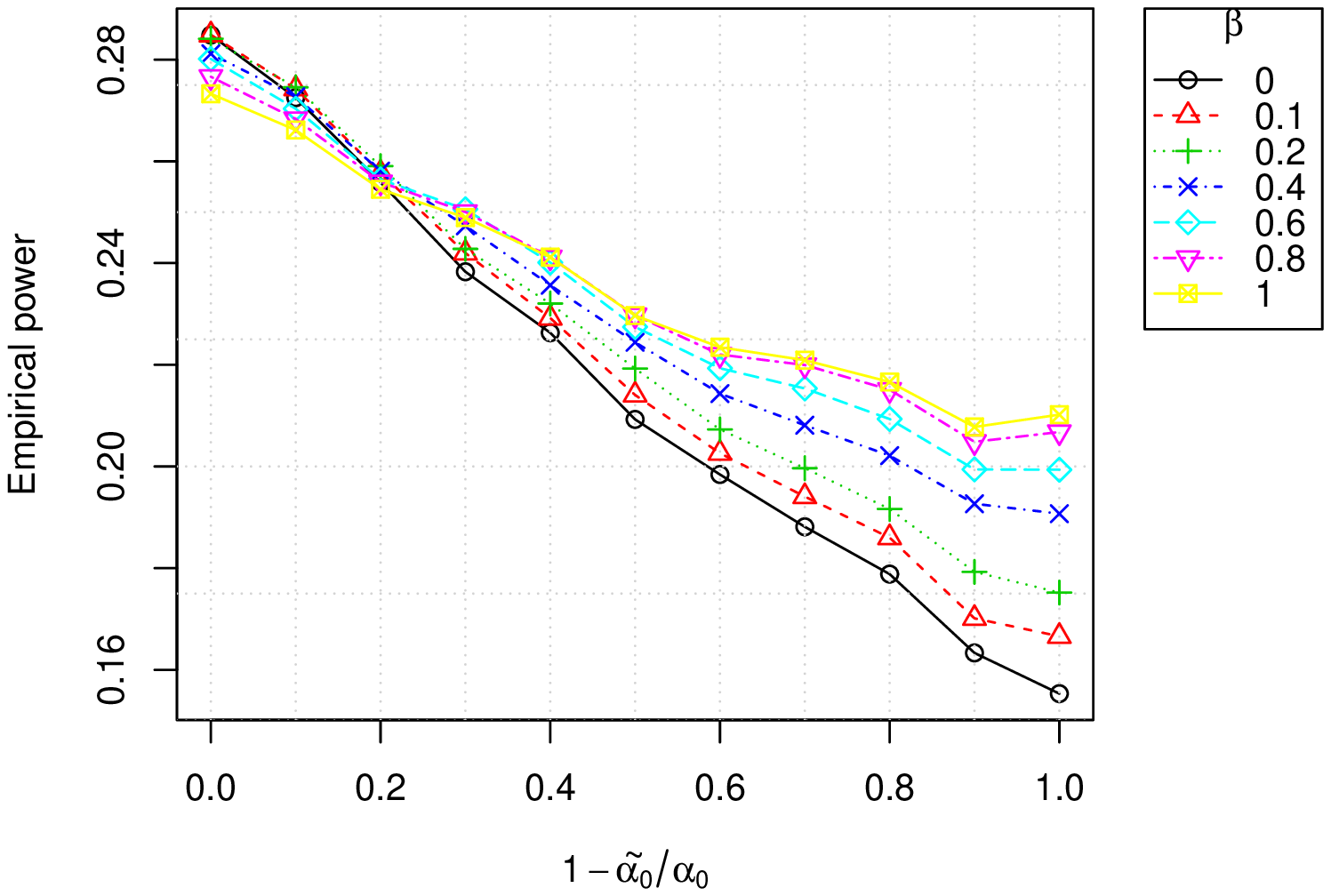}}
\end{tabular}
\caption{Simulated levels (left) and powers (right) with an $\alpha_0$-contaminated outlying cell.\label{fig:LP_2}}%
\vspace{0.3cm}
\end{figure}%

\begin{figure}[h]
  \centering
\begin{tabular}
[c]{cc}
\raisebox{-0cm}{\includegraphics[scale=0.54]
{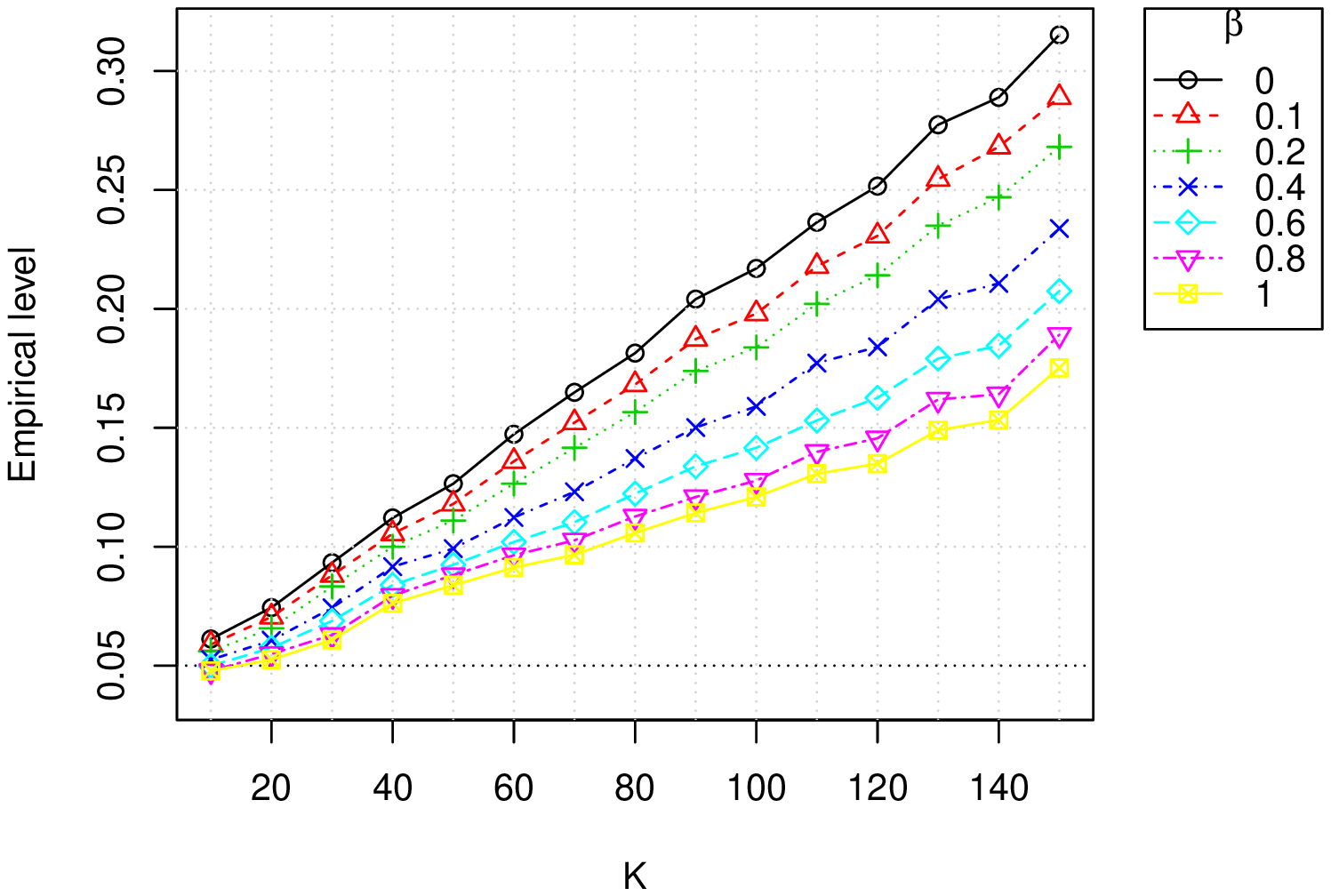}%
}%
&
\raisebox{-0cm}{\includegraphics[scale=0.54]
{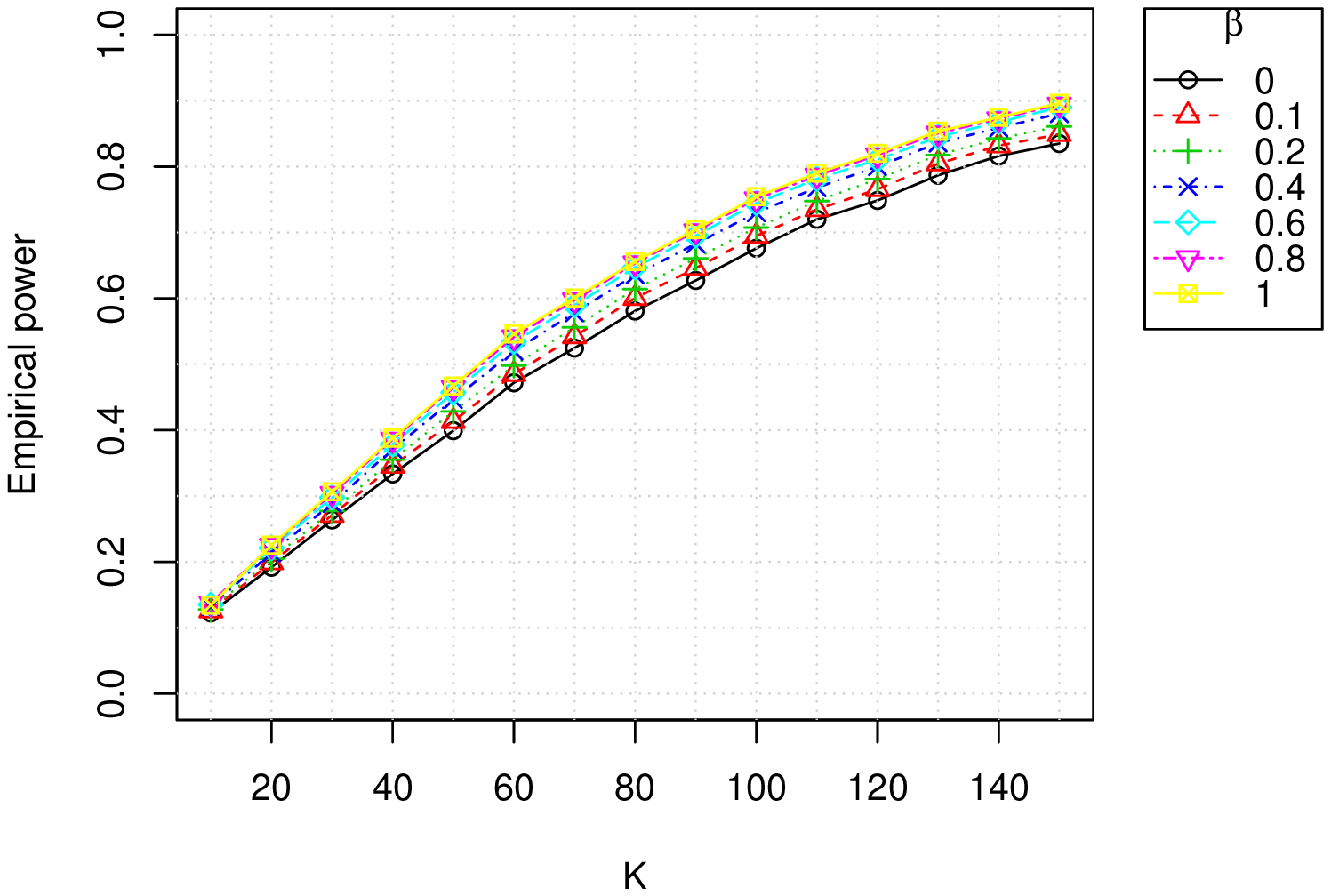}}%
\\%
\raisebox{-0cm}{\includegraphics[scale=0.54]
{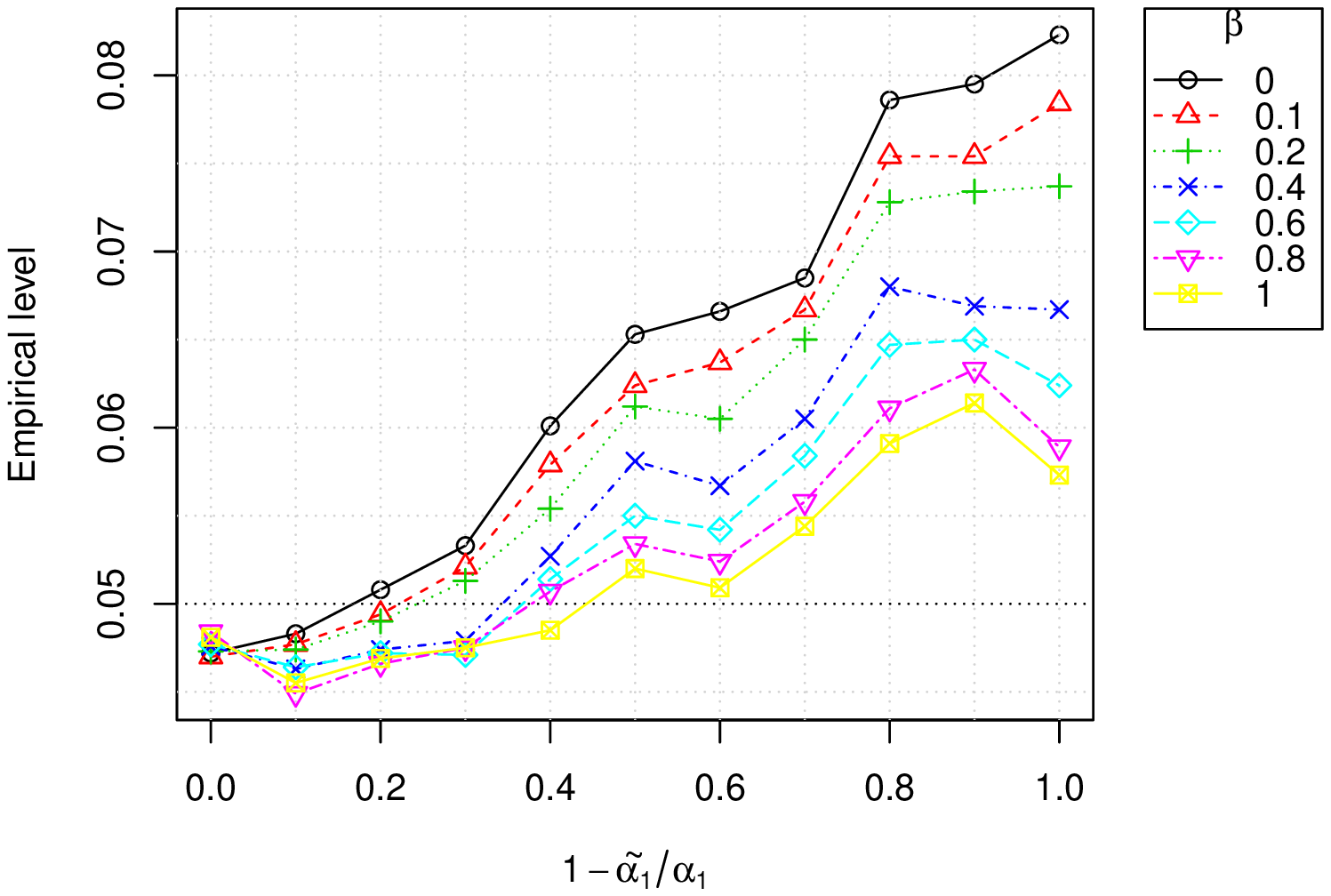}}%
&
\raisebox{-0cm}{\includegraphics[scale=0.54]
{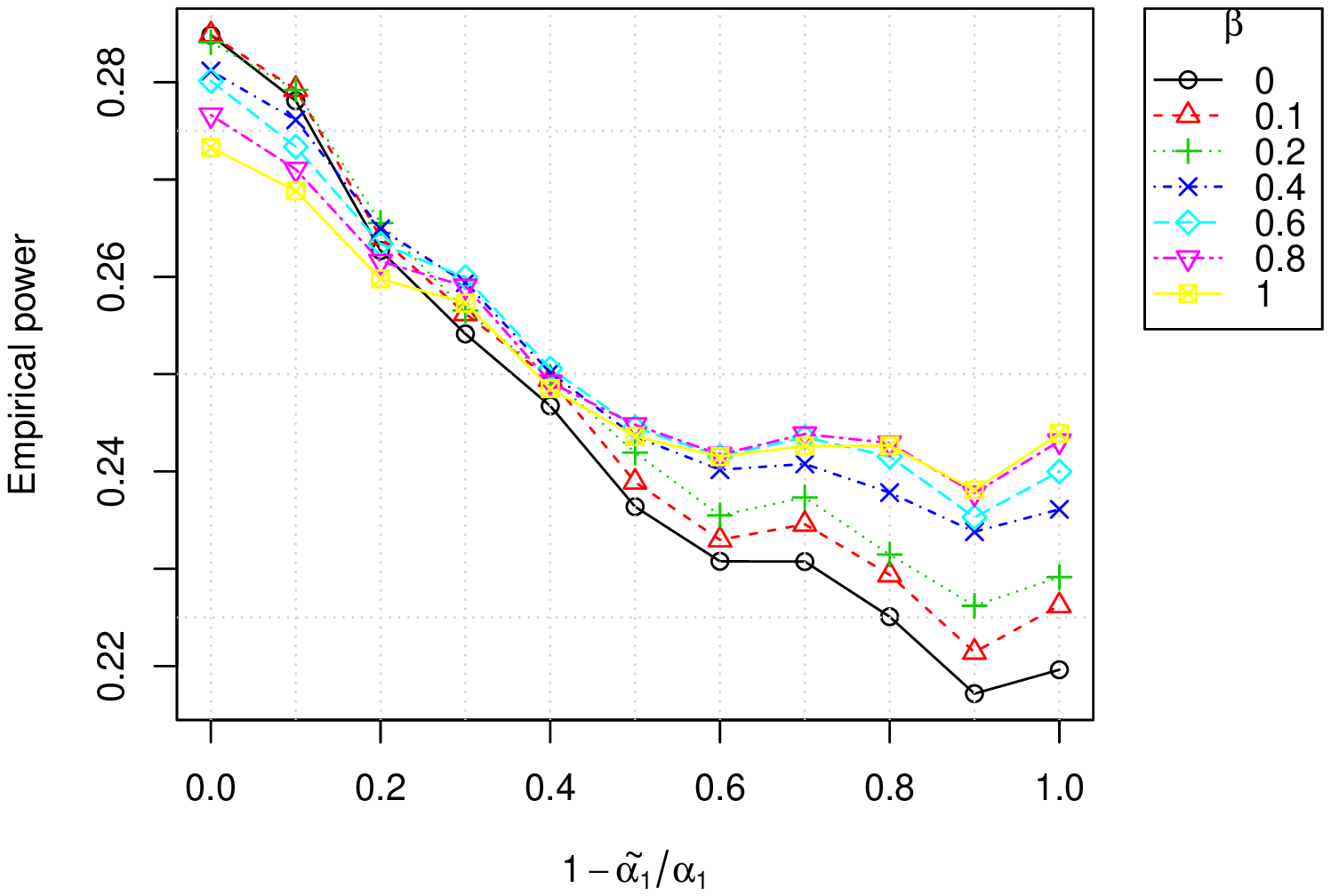}%
}%
\end{tabular}
\caption{Simulated levels (left) and powers (right) with an $\alpha_1$-contaminated outlying cell in the data.\label{fig:LP_3}}%
\end{figure}%

To evaluate the robustness of the level and  the power of the $Z$-type tests based on MDPDEs with an outlier placed on the first-row first-column cell, we perform the simulation for the same test and the same true values for the null and alternative hypotheses, in two different scenarios depending on the way the outlying cell is considered. In the first scenario, we keep $\alpha_{1}$ the same and modify the true value of $\alpha_{0}$ to be $\tilde{\alpha}_{0}\leq\alpha_{0}$, and in the second one, we keep $\alpha_{0}$ the same and modify the true value of $\alpha_{1}$ to be $\tilde{\alpha}_{1}\leq\alpha_{1}$. Both cases have been analyzed for different values of $K$ and decreasing $\tilde{\alpha}_{0}$\ in the first scenario (increasing $1-\tilde{\alpha}_{0}/\alpha_{0}$) or decreasing $\tilde{\alpha}_{1}$ in the second scenario (increasing $1-\tilde{\alpha}_{1}/\alpha_{1}$).

The results for the first scenario are presented in Figure \ref{fig:LP_2}. The empirical level for the one-shot device model with $K$ from $10$ to $150$, true value $(\alpha_{0},\alpha_{1})=(0.004,0.05)$ and $\tilde{\alpha}_{0}=0.001$ for the outlying cell is presented on the left and top panel. Similarly, the empirical power for the one-shot device model with $K$ from $10$ to $150$, true parameter $(\alpha_{0},\alpha_{1})=(0.004,0.02)$ and $\tilde{\alpha}_{0}=0.001$ for the outlying cell is presented on the right  top panel. In addition, the empirical level for the one-shot device model with $1-\tilde{\alpha}_{0}/\alpha_{0}$ from $0$ to $1$ for the outlying cell and true value $(\alpha_{0},\alpha_{1})=(0.004,0.05)$ and $K=20$ is presented on the left bottom panel. Similarly, the empirical power for the one-shot device model with $1-\tilde{\alpha}_{0}/\alpha_{0}$ from $0$ to $1$ for the outlying cell and true value and true parameter $(\alpha_{0},\alpha_{1})=(0.004,0.02)$ is presented on the right bottom panel.

Notice that the outlying cell represents  1/9 of the total observations in the last plots. For large values of $K$ (very large sample sizes, since $n=9K$), there is a large inflation in the empirical level and shrinkage of the empirical power, but for the $Z$-type test statistic based on the MDPDEs with large values of the tuning parameter $\beta$, the effect of the outlying cell is weaker in comparison to those of smaller values of $\beta$, included the MLEs ($\beta=0$). If $\tilde{\alpha}_{0}$ is separated from $\alpha_{0}$ ($1-\tilde{\alpha}_{0}/\alpha_{0}$ increases from $0$ to $1$), the empirical level of the $Z$-type test statistics based on the MDPDEs is not stable around the nominal level, being however closer as the tuning parameter $\beta$ becomes larger. If $\tilde{\alpha}_{0}$ is separated from $\alpha_{0}$ ($1-\tilde{\alpha}_{0}/\alpha_{0}$ increases from $0$ to $1$), the empirical power of the $Z$-type test statistics based on the MDPDEs decreases, being however more slowly as the tuning parameter $\beta$ becomes larger.

Figure \ref{fig:LP_3}  presents the results for the second scenario, in which $\tilde{\alpha}_{1}=0.01$ for the outlying cell on the left  top panel  and $\tilde{\alpha}_{1}=-0.01$ for the outlying cell on the right  top panel. Even though the outliers are, in the current scenario, slightly more pronunced with respect to the previous scenario, in general terms, we arrive at the same conclusions as in the previous scenario.%

The results of the tests statistics presented here show again the poor behavior in robustness of the $Z$-type tests based on the MLE of the parameters of the one-shot device model. Furthermore, the robustness properties of the $Z$-type test statistics based on the MDPDEs with large values of the tuning parameter $\beta$ are often better as they maintain both level and power in a stable manner. Moreover, the comments made at the end of Section \ref{sec6.1} for the MDPDEs regarding  moderate values of the tuning parameter $\beta$ are valid for the $Z$-type test statistics based on the MDPDEs as well.

\begin{equation}
\mathcal{T}_{1,\beta}(\boldsymbol{\alpha})=\sum\limits_{i=1}^{I}%
\sum\limits_{j=1}^{J}\left\{  \left(  \frac{F\left(  t_{j}|\lambda_{w_{i}%
}(\boldsymbol{\alpha})\right)  }{IJ}\right)  ^{1+\beta}-(1+\tfrac{1}{\beta
})\left(  \frac{F\left(  t_{j}|\lambda_{w_{i}}(\boldsymbol{\alpha})\right)
}{IJ}\right)  ^{\beta}\frac{n_{ij}}{IJK}+\tfrac{1}{\beta}\left(  \frac{n_{ij}%
}{IJK}\right)  ^{1+\beta}\right\}
\label{eq:reult5_a}
\end{equation}
\begin{equation}
\mathcal{T}_{2,\beta}(\boldsymbol{\alpha})=\sum\limits_{i=1}^{I}%
\sum\limits_{j=1}^{J}\left\{  \left(  \frac{1-F\left(  t_{j}|\lambda_{w_{i}%
}(\boldsymbol{\alpha})\right)  }{IJ}\right)  ^{1+\beta}-(1+\tfrac{1}{\beta
})\left(  \frac{1-F\left(  t_{j}|\lambda_{w_{i}}(\boldsymbol{\alpha})\right)
}{IJ}\right)  ^{\beta}\frac{K-n_{ij}}{IJK}+\tfrac{1}{\beta}\left(
\frac{K-n_{ij}}{IJK}\right)  ^{1+\beta}\right\}
\label{eq:reult5_b}
\end{equation}

\section{Concluding Remarks\label{sec7}}

In this paper, we have introduced and studied the minimum density power divergence estimators for  one-shot device testing with an accelerating factor of temperature. Based on these estimators, we have also introduced a Wald-type test statistic family. Since the maximum likelihood estimator is a particular estimator in the family of minimum density power divergence estimators developed here, the classical Wald test is also taken into account for comparison.  The results obtained in the simulation study suggest that some minimum density power divergence estimators are considerably better for the estimation of the model parameters when outliers are present in the data and at the same time not facing much loss of efficiency when outliers are not present. Similar results are obtained for some Wald-type test statistics in terms of stability with respect to  level and  power. These  proposed estimators also  give a more meaningful result in the case of ED01  tumorigenicity experiment data than the maximum likelihood estimators.

\clearpage
\appendix
\section{Proofs of Results}


\subsection{Proof of Result 1:}

We have
\begin{align*}
d_{KL}\left(  \widehat{\boldsymbol{p}},\boldsymbol{p}(\boldsymbol{\alpha
})\right)   &  =\frac{1}{IJK}\left(  s-\sum\limits_{i=1}^{I}\sum
\limits_{j=1}^{J}\log\left(  F(t_{j}|\lambda_{w_{i}}(\boldsymbol{\alpha
}))\right)  ^{n_{ij}} \right.\\
 & \left.+\sum\limits_{i=1}^{I}\sum\limits_{j=1}^{J}\log\left(
\left(  1-F(t_{j}|\lambda_{w_{i}}(\boldsymbol{\alpha}))\right)  \right)
^{K-n_{ij}}\right) \\
&  =\frac{1}{IJK}\left(  s-\log\prod\limits_{i=1}^{I}\prod\limits_{j=1}%
^{J}F(t_{j}|\lambda_{w_{i}}(\boldsymbol{\alpha}))^{n_{ij}}\right.\\
& \times \left.\left(1-F(t_{j}|\lambda_{w_{i}}(\boldsymbol{\alpha}))\right)  ^{K-n_{ij}}\right.\Bigg) \\
&  =\frac{1}{IJK}\left(  s-\log\mathcal{L}\left(  \boldsymbol{\alpha
}\left\vert K,\boldsymbol{n},\boldsymbol{t},\boldsymbol{w}\right.  \right)
\right)  ,
\end{align*}
with
\[
s=\sum\limits_{i=1}^{I}\sum\limits_{j=1}^{J}n_{ij}\log\frac{n_{ij}}{K}%
+\sum\limits_{i=1}^{I}\sum\limits_{j=1}^{J}(K-n_{ij})\log\frac{K-n_{ij}}{K},
\]
as required.


\subsection{Proof of Result 4:}
The relationship between (\ref{4b}) and $d_{\beta}\left(
\widehat{\boldsymbol{p}},\boldsymbol{p}(\boldsymbol{\alpha})\right)  $ defined
in (\ref{3}) is given by
\begin{align*}
&  \frac{1}{IJ}\sum\limits_{i=1}^{I}\sum\limits_{j=1}^{J}\left\{  \pi
_{ij}^{\beta+1}(\boldsymbol{\alpha})+(1-\pi_{ij}(\boldsymbol{\alpha}%
))^{\beta+1} \right.\\
&\left.-\frac{1+\beta}{\beta}\frac{n_{ij}}{K}\pi_{ij}^{\beta
}(\boldsymbol{\alpha})-\frac{1+\beta}{\beta}\frac{K-n_{ij}}{K}(1-\pi
_{ij}(\boldsymbol{\alpha}))^{\beta}\right\} \\
&  =(IJ)^{\beta+1}d_{\beta}\left(  \widehat{\boldsymbol{p}},\boldsymbol{p}%
(\boldsymbol{\alpha})\right)  +c,
\end{align*}
where $c$ is a constant not dependent on $\boldsymbol{\alpha}$, and so
$\widehat{\boldsymbol{\alpha}}_{\beta}$\ is the same for both cases. Hence, the
result.


\subsection{Proof of Result 5:}

We have
\begin{align}
\frac{\partial F(t_{j}|\lambda_{w_{i}}(\boldsymbol{\alpha}))}{\partial
\alpha_{0}}&=\exp\left\{  -\alpha_{0}\exp\left(  \alpha_{1}w_{i}\right)
t_{j}\right\}  \exp\left\{  \alpha_{1}w_{i}\right\}  t_{j} \nonumber  \\
&=f(t_{j}%
|\lambda_{w_{i}}(\boldsymbol{\alpha}))\frac{t_{j}}{\alpha_{0}} \label{6}%
\end{align}
and
\begin{align}
\frac{\partial F(t_{j}|\lambda_{w_{i}}(\boldsymbol{\alpha}))}{\partial
\alpha_{1}}=&\exp\left\{  -\alpha_{0}\exp\left(  \alpha_{1}w_{i}\right)
t_{j}\right\}  \nonumber \\
& \times \exp\left\{  \alpha_{1}w_{i}\right\}  \alpha_{0}t_{j}%
w_{i} \nonumber \\
=&f(t_{j}|\lambda_{w_{i}}(\boldsymbol{\alpha}))t_{j}w_{i}. \label{7}%
\end{align}
We denote
\[
d_{\beta}\left(  \widehat{\boldsymbol{p}},\boldsymbol{p}(\boldsymbol{\alpha
})\right)  =\mathcal{T}_{1,\beta}(\boldsymbol{\alpha})+\mathcal{T}_{2,\beta
}(\boldsymbol{\alpha}),
\]
where $\mathcal{T}_{1,\beta}(\boldsymbol{\alpha})$ and $\mathcal{T}_{2,\beta}(\boldsymbol{\alpha})$ are given by (\ref{eq:reult5_a}) and (\ref{eq:reult5_b}), respectively, for $\beta>0$. 

Based on (\ref{6}), we have
\begin{align*}
\frac{\partial\mathcal{T}_{1,\beta}(\boldsymbol{\alpha})}{\partial\alpha_{0}%
}=&\frac{\beta+1}{\left(  IJ\right)  ^{\beta+1}}\sum\limits_{i=1}^{I}%
\sum\limits_{j=1}^{J}\left(  F(t_{j}|\lambda_{w_{i}}(\boldsymbol{\alpha
}))-\frac{n_{ij}}{K}\right)\\
& \times  f(t_{j}|\lambda_{w_{i}}(\boldsymbol{\alpha
}))\frac{t_{j}}{\alpha_{0}}F^{^{\beta-1}}(t_{j}|\lambda_{w_{i}}%
(\boldsymbol{\alpha}))
\end{align*}
and
\begin{align*}
\frac{\partial\mathcal{T}_{2,\beta}(\boldsymbol{\alpha})}{\partial\alpha_{0}%
}=&\frac{\beta+1}{\left(  IJ\right)  ^{\beta+1}}\sum\limits_{i=1}^{I}%
\sum\limits_{j=1}^{J}\left(  F(t_{j}|\lambda_{w_{i}}(\boldsymbol{\alpha
}))-\frac{n_{ij}}{K}\right)\\
&\times  f(t_{j}|\lambda_{w_{i}}(\boldsymbol{\alpha
}))\frac{t_{j}}{\alpha_{0}}\left(  1-F(t_{j}|\lambda_{w_{i}}%
(\boldsymbol{\alpha}))\right)  ^{^{\beta-1}}.
\end{align*}
On the other hand, by (\ref{7}), we have
\begin{align*}
\frac{\partial\mathcal{T}_{1,\beta}(\boldsymbol{\alpha})}{\partial\alpha_{1}%
}=&\frac{\beta+1}{\left(  IJ\right)  ^{\beta+1}}\sum\limits_{i=1}^{I}%
\sum\limits_{j=1}^{J}\left(  F(t_{j}|\lambda_{w_{i}}(\boldsymbol{\alpha
}))-\frac{n_{ij}}{K}\right)\\
&\times  f(t_{j}|\lambda_{w_{i}}(\boldsymbol{\alpha
}))t_{j}w_{i}F^{^{\beta-1}}(t_{j}|\lambda_{w_{i}}(\boldsymbol{\alpha}))
\end{align*}
and
\begin{align*}
\frac{\partial\mathcal{T}_{2,\beta}(\boldsymbol{\alpha})}{\partial\alpha_{1}%
}=&\frac{\beta+1}{\left(  IJ\right)  ^{\beta+1}}\sum\limits_{i=1}^{I}%
\sum\limits_{j=1}^{J}\left(  F(t_{j}|\lambda_{w_{i}}(\boldsymbol{\alpha
}))-\frac{n_{ij}}{K}\right) \\
&\times f(t_{j}|\lambda_{w_{i}}(\boldsymbol{\alpha
}))t_{j}w_{i}\left(  1-F(t_{j}|\lambda_{w_{i}}(\boldsymbol{\alpha}))\right)
^{^{\beta-1}}.
\end{align*}
Finally, the system of equations is given by
\begin{align*}
\frac{\left(  IJ\right)  ^{\beta+1}}{\beta+1}\left(  \frac{\partial
\mathcal{T}_{1,\beta}(\boldsymbol{\alpha})}{\partial\alpha_{0}}+\frac
{\partial\mathcal{T}_{2,\beta}(\boldsymbol{\alpha})}{\partial\alpha_{0}%
}\right)   &  =0,\\
\frac{\left(  IJ\right)  ^{\beta+1}}{\beta+1}\left(  \frac{\partial
\mathcal{T}_{1,\beta}(\boldsymbol{\alpha})}{\partial\alpha_{1}}+\frac
{\partial\mathcal{T}_{2,\beta}(\boldsymbol{\alpha})}{\partial\alpha_{1}%
}\right)   &  =0.
\end{align*}
If we consider $\beta=0$ in (\ref{5}) and (\ref{5B}), we get the system needed
to solve in order to get the maximum likelihood estimator (MLE). Hence, the
previous system of equations is valid not only for tuning parameters $\beta
>0$, but also for $\beta=0$.


\subsection{Proof of Result 6:}
Based on Ghosh and Basu (2013) and also on Definition \ref{def1}, we have%
\[
\sqrt{IJK}\left(  \widehat{\boldsymbol{\alpha}}_{\beta}-\boldsymbol{\alpha
}_{0}\right)  \overset{\mathcal{L}}{\underset{IJK\mathcal{\rightarrow}%
\infty}{\longrightarrow}}\mathcal{N}\left(  \boldsymbol{0},\boldsymbol{J}%
_{\beta}^{-1}(\boldsymbol{\alpha}_{0})\boldsymbol{K}_{\beta}%
(\boldsymbol{\alpha}_{0})\boldsymbol{J}_{\beta}^{-1}(\boldsymbol{\alpha}%
_{0})\right)  ,
\]
where
\begin{align*}
\boldsymbol{J}_{\beta}(\boldsymbol{\alpha})   = &\frac{1}{IJK}\sum
\limits_{i=1}^{I}\sum\limits_{j=1}^{J}\sum\limits_{k=1}^{K}\boldsymbol{J}%
_{ij,\beta}(\boldsymbol{\alpha})\\
=&\frac{1}{IJ}\sum\limits_{i=1}^{I}\sum\limits_{j=1}^{J}\boldsymbol{J}_{ij,\beta}(\boldsymbol{\alpha}),\\
\boldsymbol{J}_{ij,\beta}(\boldsymbol{\alpha})    =&\boldsymbol{u}%
_{ij}(\boldsymbol{\alpha})\boldsymbol{u}_{ij}^{T}(\boldsymbol{\alpha}%
)F^{\beta+1}(t_{j}|\lambda_{w_{i}}(\boldsymbol{\alpha}))\\
&+\boldsymbol{v}%
_{ij}(\boldsymbol{\alpha})\boldsymbol{v}_{ij}^{T}(\boldsymbol{\alpha
})(1-F(t_{j}|\lambda_{w_{i}}(\boldsymbol{\alpha})))^{\beta+1}\\
  = &t_{j}^{2}f^{2}(t_{j}|\lambda_{w_{i}}(\boldsymbol{\alpha}))%
\begin{pmatrix}
\frac{1}{\alpha_{0}^{2}} & \frac{w_{i}}{\alpha_{0}}\\
\frac{w_{i}}{\alpha_{0}} & w_{i}^{2}%
\end{pmatrix}
\left[  F^{\beta-1}(t_{j}|\lambda_{w_{i}}(\boldsymbol{\alpha})) \right.\\
& \left. +(1-F(t_{j}%
|\lambda_{w_{i}}(\boldsymbol{\alpha})))^{\beta-1}\right],
\end{align*}

\begin{align*}
\boldsymbol{u}_{ij}(\boldsymbol{\alpha})   = &  \frac{\partial\log
F(t_{j}|\lambda_{w_{i}}(\boldsymbol{\alpha}))}{\partial\boldsymbol{\alpha}%
}\\
=&\frac{1}{F(t_{j}|\lambda_{w_{i}}(\boldsymbol{\alpha}))}\frac{\partial
}{\partial\boldsymbol{\alpha}}F(t_{j}|\lambda_{w_{i}}(\boldsymbol{\alpha})),\\
\boldsymbol{v}_{ij}(\boldsymbol{\alpha})   =   & \frac{\partial\log\left[
1-F(t_{j}|\lambda_{w_{i}}(\boldsymbol{\alpha}))\right]  }{\partial
\boldsymbol{\alpha}}\\
=&-\frac{1}{1-F(t_{j}|\lambda_{w_{i}}(\boldsymbol{\alpha
}))}\frac{\partial}{\partial\boldsymbol{\alpha}}F(t_{j}|\lambda_{w_{i}%
}(\boldsymbol{\alpha})),\\
\frac{\partial}{\partial\boldsymbol{\alpha}}F(t_{j}|\lambda_{w_{i}%
}(\boldsymbol{\alpha}))    = & -\frac{\partial}{\partial\boldsymbol{\alpha}}%
\exp\left\{  -\alpha_{0}\exp\left\{  \alpha_{1}w_{i}\right\}  t_{j}\right\}\\
  =&
\begin{pmatrix}
\frac{1}{\alpha_{0}}\\
w_{i}%
\end{pmatrix}
t_{j}f(t_{j}|\lambda_{w_{i}}(\boldsymbol{\alpha})),
\end{align*}
and%
\begin{align*}
\boldsymbol{K}_{\beta}(\boldsymbol{\alpha})    = &\frac{1}{IJK}\sum
\limits_{i=1}^{I}\sum\limits_{j=1}^{J}\sum\limits_{k=1}^{K}\boldsymbol{K}%
_{ij,\beta}(\boldsymbol{\alpha})\\
=&\frac{1}{IJ}\sum\limits_{i=1}^{I}%
\sum\limits_{j=1}^{J}\boldsymbol{K}_{ij,\beta}(\boldsymbol{\alpha}),\\
\boldsymbol{K}_{ij,\beta}(\boldsymbol{\alpha})   =&   \boldsymbol{S}%
_{ij,\boldsymbol{\beta}}(\boldsymbol{\alpha})-\boldsymbol{\xi}_{ij,\beta
}(\boldsymbol{\alpha})\boldsymbol{\xi}_{ij,\beta}^{T}(\boldsymbol{\alpha}),\\
\boldsymbol{S}_{ij,\boldsymbol{\beta}}(\boldsymbol{\alpha})  
= &\boldsymbol{u}_{ij}(\boldsymbol{\alpha})\boldsymbol{u}_{ij}^{T}%
(\boldsymbol{\alpha})F^{2\beta+1}(t_{j}|\lambda_{w_{i}}(\boldsymbol{\alpha
}))\\
&+\boldsymbol{v}_{ij}(\boldsymbol{\alpha})\boldsymbol{v}_{ij}^{T}%
(\boldsymbol{\alpha})(1-F(t_{j}|\lambda_{w_{i}}(\boldsymbol{\alpha}%
)))^{2\beta+1}\\
  = & t_{j}^{2}f^{2}(t_{j}|\lambda_{w_{i}}(\boldsymbol{\alpha})))%
\begin{pmatrix}
\frac{1}{\alpha_{0}^{2}} & \frac{w_{i}}{\alpha_{0}}\\
\frac{w_{i}}{\alpha_{0}} & w_{i}^{2}%
\end{pmatrix}
\left[  F^{2\beta-1}(t_{j}|\lambda_{w_{i}}(\boldsymbol{\alpha})) \right.\\
&\left.+(1-F(t_{j}%
|\lambda_{w_{i}}(\boldsymbol{\alpha})))^{2\beta-1}\right]  ,\\
\boldsymbol{\xi}_{ij,\beta}(\boldsymbol{\alpha})    =&\boldsymbol{u}%
_{ij}(\boldsymbol{\alpha})F^{\beta+1}(t_{j}|\lambda_{w_{i}}(\boldsymbol{\alpha
}))\\
&+\boldsymbol{v}_{ij}(\boldsymbol{\alpha})(1-F(t_{j}|\lambda_{w_{i}%
}(\boldsymbol{\alpha})))^{\beta+1}\\
  =&
\begin{pmatrix}
\frac{1}{\alpha_{0}}\\
w_{i}%
\end{pmatrix}
t_{j}f(t_{j}|\lambda_{w_{i}}(\boldsymbol{\alpha}))\left[  F^{\beta}%
(t_{j}|\lambda_{w_{i}}(\boldsymbol{\alpha}))\right.\\
&\left.-(1-F(t_{j}|\lambda_{w_{i}%
}(\boldsymbol{\alpha})))^{\beta}\right]  .
\end{align*}
Since $I$, $J$ are fixed and $IJK\mathcal{\rightarrow}\infty$, it follows that
$K\mathcal{\rightarrow}\infty$ and%
\[
\sqrt{K}\left(  \widehat{\boldsymbol{\alpha}}_{\beta}-\boldsymbol{\alpha}%
_{0}\right)   \overset{\mathcal{L}%
}{\underset{K\mathcal{\rightarrow}\infty}{\longrightarrow}}\mathcal{N}\left(
\boldsymbol{0},\boldsymbol{\bar{J}}_{\beta}^{-1}(\boldsymbol{\alpha}%
_{0})\boldsymbol{\bar{K}}_{\beta}(\boldsymbol{\alpha}_{0})\boldsymbol{\bar{J}%
}_{\beta}^{-1}(\boldsymbol{\alpha}_{0})\right)
\]
where
\begin{align*}
\boldsymbol{\bar{J}}_{\beta}^{-1}(\boldsymbol{\alpha}_{0})\boldsymbol{\bar{K}%
}_{\beta}(\boldsymbol{\alpha}_{0})\boldsymbol{\bar{J}}_{\beta}^{-1}%
(\boldsymbol{\alpha}_{0})  &  =\frac{1}{IJ}\boldsymbol{J}_{\beta}%
^{-1}(\boldsymbol{\alpha}_{0})\boldsymbol{K}_{\beta}(\boldsymbol{\alpha}%
_{0})\boldsymbol{J}_{\beta}^{-1}(\boldsymbol{\alpha}_{0}),\\
\boldsymbol{\bar{J}}_{\beta}(\boldsymbol{\alpha}_{0})  &  =(IJ)\boldsymbol{J}%
_{\beta}(\boldsymbol{\alpha}_{0}),\\
\boldsymbol{\bar{K}}_{\beta}(\boldsymbol{\alpha}_{0})  &  =(IJ)\boldsymbol{K}%
_{\beta}(\boldsymbol{\alpha}_{0}).
\end{align*}

\subsection{Proof of Result 7:}

The Fisher information matrix for $IJK$\ observations is
\[
\boldsymbol{I}_{IJK,F}\left(  \boldsymbol{\alpha}\right)  =E\left[
-\frac{\partial\boldsymbol{v}^{T}\left(  \boldsymbol{\alpha}\left\vert
K,\boldsymbol{n},\boldsymbol{t},\boldsymbol{w}\right.  \right)  }%
{\partial\boldsymbol{\alpha}}\right]  ,
\]
where%
\[
\boldsymbol{v}\left(  \boldsymbol{\alpha}\left\vert K,\boldsymbol{n}%
,\boldsymbol{t},\boldsymbol{w}\right.  \right)  =\frac{\partial\log
\mathcal{L}\left(  \boldsymbol{\alpha}\left\vert K,\boldsymbol{n}%
,\boldsymbol{t},\boldsymbol{w}\right.  \right)  }{\partial\boldsymbol{\alpha}%
}.
\]
From (\ref{1}),
\begin{align*}
\boldsymbol{I}_{IJK,F}\left(  \boldsymbol{\alpha}\right) & =IJK\;E\left[
\frac{\partial^{2}d_{KL}\left(  \widehat{\boldsymbol{p}},\boldsymbol{p}%
(\boldsymbol{\alpha})\right)  }{\partial\boldsymbol{\alpha}\partial
\boldsymbol{\alpha}^{T}}\right] \\
& =IJK\;E\left[  \frac{\partial\boldsymbol{u}%
^{T}\left(  \boldsymbol{\alpha}\left\vert K,\boldsymbol{n},\boldsymbol{t}%
,\boldsymbol{w}\right.  \right)  }{\partial\boldsymbol{\alpha}}\right]  ,
\end{align*}
where%
\begin{align*}
\boldsymbol{u}\left(  \boldsymbol{\alpha}\left\vert K,\boldsymbol{n}%
,\boldsymbol{t},\boldsymbol{w}\right.  \right)  &=\frac{\partial d_{KL}\left(
\widehat{\boldsymbol{p}},\boldsymbol{p}(\boldsymbol{\alpha})\right)
}{\partial\boldsymbol{\alpha}}\\
&=\frac{\partial\mathcal{T}_{1,\beta
=0}(\boldsymbol{\alpha})}{\partial\boldsymbol{\alpha}}+\frac{\partial
\mathcal{T}_{2,\beta=0}(\boldsymbol{\alpha})}{\partial\boldsymbol{\alpha}}.
\end{align*}
The Fisher information matrix for a single observation, i.e., the Fisher information matrix for the one-shot device model is%
\[
\boldsymbol{I}_{F}\left(  \boldsymbol{\alpha}\right)  =\frac{1}{IJK}%
\boldsymbol{I}_{IJK,F}\left(  \boldsymbol{\alpha}\right)  =E\left[
\frac{\partial\boldsymbol{u}^{T}\left(  \boldsymbol{\alpha}\left\vert
K,\boldsymbol{n},\boldsymbol{t},\boldsymbol{w}\right.  \right)  }%
{\partial\boldsymbol{\alpha}}\right]  .
\]
From Result \ref{Th5}, the first and second components of $\boldsymbol{u}%
\left(  \boldsymbol{\alpha}\left\vert K,\boldsymbol{n},\boldsymbol{t}%
,\boldsymbol{w}\right.  \right)  $ are 
\begin{small}
\begin{align*}
&  u_{1}\left(  \boldsymbol{\alpha}\left\vert K,\boldsymbol{n},\boldsymbol{t}%
,\boldsymbol{w}\right.  \right)  =\frac{\partial\mathcal{T}_{1,\beta
=0}(\boldsymbol{\alpha})}{\partial\alpha_{0}}+\frac{\partial\mathcal{T}%
_{2,\beta=0}(\boldsymbol{\alpha})}{\partial\alpha_{0}}\\
&  =\frac{1}{IJK}\sum\limits_{i=1}^{I}\sum\limits_{j=1}^{J}\left(
K\ F(t_{j}|\lambda_{w_{i}}(\boldsymbol{\alpha}))-n_{ij}\right)\\
& \ \ \times  f(t_{j}%
|\lambda_{w_{i}}(\boldsymbol{\alpha}))\frac{t_{j}}{\alpha_{0}}\left[
F^{-1}(t_{j}|\lambda_{w_{i}}(\boldsymbol{\alpha}))+\left(  1-F(t_{j}%
|\lambda_{w_{i}}(\boldsymbol{\alpha}))\right)  ^{-1}\right] \\
&  =\frac{1}{IJK}\sum\limits_{i=1}^{I}\sum\limits_{j=1}^{J}\frac
{K\ F(t_{j}|\lambda_{w_{i}}(\boldsymbol{\alpha}))-n_{ij}}{F(t_{j}%
|\lambda_{w_{i}}(\boldsymbol{\alpha}))\left(  1-F(t_{j}|\lambda_{w_{i}%
}(\boldsymbol{\alpha}))\right)  }f(t_{j}|\lambda_{w_{i}}(\boldsymbol{\alpha
}))\frac{t_{j}}{\alpha_{0}}%
\end{align*}
\end{small}
and%
\begin{small}
\begin{align*}
&  u_{2}\left(  \boldsymbol{\alpha}\left\vert K,\boldsymbol{n},\boldsymbol{t}%
,\boldsymbol{w}\right.  \right)  =\frac{\partial\mathcal{T}_{1,\beta
=0}(\boldsymbol{\alpha})}{\partial\alpha_{1}}+\frac{\partial\mathcal{T}%
_{2,\beta=0}(\boldsymbol{\alpha})}{\partial\alpha_{1}}\\
&  =\frac{1}{IJK}\sum\limits_{i=1}^{I}\sum\limits_{j=1}^{J}\left(
K\ F(t_{j}|\lambda_{w_{i}}(\boldsymbol{\alpha}))-n_{ij}\right)\\
& \ \ \times   f(t_{j}%
|\lambda_{w_{i}}(\boldsymbol{\alpha}))t_{j}w_{i}\left[  F^{-1}(t_{j}%
|\lambda_{w_{i}}(\boldsymbol{\alpha}))+\left(  1-F(t_{j}|\lambda_{w_{i}%
}(\boldsymbol{\alpha}))\right)  ^{-1}\right] \\
&  =\frac{1}{IJK}\sum\limits_{i=1}^{I}\sum\limits_{j=1}^{J}\frac
{K\ F(t_{j}|\lambda_{w_{i}}(\boldsymbol{\alpha}))-n_{ij}}{F(t_{j}%
|\lambda_{w_{i}}(\boldsymbol{\alpha}))\left(  1-F(t_{j}|\lambda_{w_{i}%
}(\boldsymbol{\alpha}))\right)  }f(t_{j}|\lambda_{w_{i}}(\boldsymbol{\alpha
}))t_{j}w_{i},
\end{align*}
\end{small}
respectively. The $(1,1)$th term of $\boldsymbol{I}_{F}\left(  \boldsymbol{\alpha}\right)  $ is
the expectation of%
\begin{footnotesize}
\begin{align*}
&\frac{\partial u_{1}\left(  \boldsymbol{\alpha}\left\vert K,\boldsymbol{n}%
,\boldsymbol{t},\boldsymbol{w}\right.  \right)  }{\partial\alpha_{0}} \\
 &
=\frac{1}{IJK}\sum\limits_{i=1}^{I}\sum\limits_{j=1}^{J}\left\{  -\frac{t_{j}%
}{\alpha_{0}^{2}}\frac{K\ F(t_{j}|\lambda_{w_{i}}(\boldsymbol{\alpha}%
))-n_{ij}}{F(t_{j}|\lambda_{w_{i}}(\boldsymbol{\alpha}))\left(  1-F(t_{j}%
|\lambda_{w_{i}}(\boldsymbol{\alpha}))\right)  }f(t_{j}|\lambda_{w_{i}%
}(\boldsymbol{\alpha}))\right. \\
& \ \ +\left.  \frac{\partial f(t_{j}|\lambda_{w_{i}}(\boldsymbol{\alpha}%
))}{\partial\alpha_{0}}\frac{t_{j}}{\alpha_{0}}\frac{K\ F(t_{j}|\lambda
_{w_{i}}(\boldsymbol{\alpha}))-n_{ij}}{F(t_{j}|\lambda_{w_{i}}%
(\boldsymbol{\alpha}))\left(  1-F(t_{j}|\lambda_{w_{i}}(\boldsymbol{\alpha
}))\right)  }\right. \\
&  \ \ +\left.  \frac{\partial}{\partial\alpha_{0}}\left(  \frac{K\ F(t_{j}%
|\lambda_{w_{i}}(\boldsymbol{\alpha}))-n_{ij}}{F(t_{j}|\lambda_{w_{i}%
}(\boldsymbol{\alpha}))\left(  1-F(t_{j}|\lambda_{w_{i}}(\boldsymbol{\alpha
}))\right)  }\right)  \frac{t_{j}}{\alpha_{0}}f(t_{j}|\lambda_{w_{i}%
}(\boldsymbol{\alpha}))\right\}  .
\end{align*}
\end{footnotesize}
Since the expectation of the first two summands of $\partial u_{1}\left(
\boldsymbol{\alpha}\left\vert K,\boldsymbol{n},\boldsymbol{t},\boldsymbol{w}%
\right.  \right)  /\partial\alpha_{0}$\ are zero, the interest is  on
the expectation of $L_{ij} $ which is given in (\ref{eq:result7}).

\begin{align}
L_{ij}  &  =\frac{\partial}{\partial\alpha_{0}}\left(  \frac{K\ F(t_{j}%
|\lambda_{w_{i}}(\boldsymbol{\alpha}))-n_{ij}}{F(t_{j}|\lambda_{w_{i}%
}(\boldsymbol{\alpha}))\left(  1-F(t_{j}|\lambda_{w_{i}}(\boldsymbol{\alpha
}))\right)  }\right)  \frac{t_{j}}{\alpha_{0}}f(t_{j}|\lambda_{w_{i}%
}(\boldsymbol{\alpha}))\nonumber \\
&  =\frac{K\frac{t_{j}}{\alpha_{0}}f(t_{j}|\lambda_{w_{i}}(\boldsymbol{\alpha
}))F(t_{j}|\lambda_{w_{i}}(\boldsymbol{\alpha}))\left(  1-F(t_{j}%
|\lambda_{w_{i}}(\boldsymbol{\alpha}))\right)  }{F(t_{j}|\lambda_{w_{i}%
}(\boldsymbol{\alpha}))^{2}\left(  1-F(t_{j}|\lambda_{w_{i}}%
(\boldsymbol{\alpha}))\right)  ^{2}}\frac{t_{j}}{\alpha_{0}}f(t_{j}%
|\lambda_{w_{i}}(\boldsymbol{\alpha})) \nonumber \\
&  -\frac{\frac{\partial}{\partial\alpha_{0}}\left[  F(t_{j}|\lambda_{w_{i}%
}(\boldsymbol{\alpha}))\left(  1-F(t_{j}|\lambda_{w_{i}}(\boldsymbol{\alpha
}))\right)  \right]  \left(  K\ F(t_{j}|\lambda_{w_{i}}(\boldsymbol{\alpha
}))-n_{ij}\right)  }{F(t_{j}|\lambda_{w_{i}}(\boldsymbol{\alpha}))^{2}\left(
1-F(t_{j}|\lambda_{w_{i}}(\boldsymbol{\alpha}))\right)  ^{2}}\frac{t_{j}%
}{\alpha_{0}}f(t_{j}|\lambda_{w_{i}}(\boldsymbol{\alpha})).
\label{eq:result7}
\end{align}

The expectation of the second summand of $L_{ij}$ is zero and hence%
\[
E[L_{ij}]=\frac{K\left(  \frac{t_{j}}{\alpha_{0}}\right)  ^{2}f^{2}%
(t_{j}|\lambda_{w_{i}}(\boldsymbol{\alpha}))}{F(t_{j}|\lambda_{w_{i}%
}(\boldsymbol{\alpha}))\left(  1-F(t_{j}|\lambda_{w_{i}}(\boldsymbol{\alpha
}))\right)  }.
\]
These finally yield the $(1,1)$th term of $\boldsymbol{I}_{F}\left(  \boldsymbol{\alpha
}\right)  $ as
\begin{align*}
&E\left[  \frac{\partial u_{1}\left(  \boldsymbol{\alpha}\left\vert
K,\boldsymbol{n},\boldsymbol{t},\boldsymbol{w}\right.  \right)  }%
{\partial\alpha_{0}}\right]  \\ &  =\frac{K}{IJK}\sum\limits_{i=1}^{I}%
\sum\limits_{j=1}^{J}\frac{\left(  \frac{t_{j}}{\alpha_{0}}\right)  ^{2}%
f^{2}(t_{j}|\lambda_{w_{i}}(\boldsymbol{\alpha}))}{F(t_{j}|\lambda_{w_{i}%
}(\boldsymbol{\alpha}))\left(  1-F(t_{j}|\lambda_{w_{i}}(\boldsymbol{\alpha
}))\right)  }\\
&  =\frac{1}{IJ}\sum\limits_{i=1}^{I}\sum\limits_{j=1}^{J}\frac{\left(
\frac{t_{j}}{\alpha_{0}}\right)  ^{2}f^{2}(t_{j}|\lambda_{w_{i}}%
(\boldsymbol{\alpha}))}{F(t_{j}|\lambda_{w_{i}}(\boldsymbol{\alpha}))\left(
1-F(t_{j}|\lambda_{w_{i}}(\boldsymbol{\alpha}))\right)  }.
\end{align*}
The rest of the terms of $\boldsymbol{I}_{F}\left(  \boldsymbol{\alpha
}\right)  $ can be obtained in a similar manner. On the other hand, from Theorem
\ref{Th5}, sustituting $\beta=0$ into $\boldsymbol{J}_{\beta}(\boldsymbol{\alpha
})=\frac{1}{IJ}\boldsymbol{\bar{J}}_{\beta}(\boldsymbol{\alpha})=$ and
$\boldsymbol{K}_{\beta}(\boldsymbol{\alpha})=\frac{1}{IJ}\boldsymbol{\bar{K}%
}_{\beta}(\boldsymbol{\alpha})$, we simply obtain $\boldsymbol{J}_{\beta
=0}(\boldsymbol{\alpha})=\boldsymbol{K}_{\beta=0}(\boldsymbol{\alpha
})=\boldsymbol{I}_{F}(\boldsymbol{\alpha})$.


\subsection{Proof of Result 9:}
Let $\boldsymbol{\alpha}_{0}$ be the true value of parameter
$\boldsymbol{\alpha}.$ It is clear that under (\ref{W1})
\[
\boldsymbol{m}^{T}\widehat{\boldsymbol{\alpha}}_{\beta}-d=\boldsymbol{m}%
^{T}(\widehat{\boldsymbol{\alpha}}_{\beta}-\boldsymbol{\alpha}_{0})
\]
and we know
\[
\sqrt{K}(\widehat{\boldsymbol{\alpha}}_{\beta}-\boldsymbol{\alpha}%
_{0})\overset{\mathcal{L}}{\underset{K\mathcal{\rightarrow}\infty
}{\longrightarrow}}\mathcal{N}(\boldsymbol{0},\boldsymbol{\bar{J}}_{\beta
}^{-1}(\boldsymbol{\alpha}_{0})\boldsymbol{\bar{K}}_{\beta}(\boldsymbol{\alpha
}_{0})\boldsymbol{\bar{J}}_{\beta}^{-1}(\boldsymbol{\alpha}_{0})),
\]
from which it follows that
\[
\sqrt{K}(\boldsymbol{m}^{T}\widehat{\boldsymbol{\alpha}}_{\beta}%
-d)\overset{\mathcal{L}}{\underset{K\mathcal{\rightarrow}\infty
}{\longrightarrow}}\mathcal{N}(0,\boldsymbol{m}^{T}\boldsymbol{\bar{J}}%
_{\beta}^{-1}(\boldsymbol{\alpha}_{0})\boldsymbol{\bar{K}}_{\beta
}(\boldsymbol{\alpha}_{0})\boldsymbol{\bar{J}}_{\beta}^{-1}(\boldsymbol{\alpha
}_{0})\boldsymbol{m}).
\]
Dividing the left hand side by $$\sqrt{\boldsymbol{m}^{T}\boldsymbol{\bar{J}%
}_{\beta}^{-1}(\widehat{\boldsymbol{\alpha}}_{\beta})\boldsymbol{\bar{K}%
}_{\beta}(\widehat{\boldsymbol{\alpha}}_{\beta})\boldsymbol{\bar{J}}_{\beta
}^{-1}(\widehat{\boldsymbol{\alpha}}_{\beta})\boldsymbol{m}},$$ since
$\boldsymbol{m}^{T}\boldsymbol{\bar{J}}_{\beta}^{-1}%
(\widehat{\boldsymbol{\alpha}}_{\beta})\boldsymbol{\bar{K}}_{\beta
}(\widehat{\boldsymbol{\alpha}}_{\beta})\boldsymbol{\bar{J}}_{\beta}%
^{-1}(\widehat{\boldsymbol{\alpha}}_{\beta})\boldsymbol{m}$ is a consistent
estimator of $\boldsymbol{m}^{T}\boldsymbol{\bar{J}}_{\beta}^{-1}%
(\boldsymbol{\alpha}_{0})\boldsymbol{\bar{K}}_{\beta}(\boldsymbol{\alpha}%
_{0})\boldsymbol{\bar{J}}_{\beta}^{-1}(\boldsymbol{\alpha}_{0})\boldsymbol{m}%
$, the desired result is obtained.

\subsection{Proof of Result 10:}

The power function at $\boldsymbol{\alpha}^{\ast}$ of $Z_{K}%
(\widehat{\boldsymbol{\alpha}}_{\beta})$ is given by equation (\ref{eq:reuslt10}).

\begin{align}
\pi\left(  \boldsymbol{\alpha}^{\ast}\right)   &  =\Pr\left(  \left\vert
Z_{K}(\widehat{\boldsymbol{\alpha}}_{\beta})\right\vert >z_{\frac{\alpha}{2}%
}|\boldsymbol{\alpha}=\boldsymbol{\alpha}^{\ast}\right) \nonumber \\
&  =2\Pr\left(  Z_{K}(\widehat{\boldsymbol{\alpha}}_{\beta})>z_{\frac{\alpha
}{2}}|\boldsymbol{\alpha}=\boldsymbol{\alpha}^{\ast}\right) \nonumber \\
&  =2\Pr\left(  \sqrt{\frac{K}{\boldsymbol{m}^{T}\boldsymbol{\bar{J}}_{\beta
}^{-1}(\widehat{\boldsymbol{\alpha}}_{\beta})\boldsymbol{\bar{K}}_{\beta
}(\widehat{\boldsymbol{\alpha}}_{\beta})\boldsymbol{\bar{J}}_{\beta}%
^{-1}(\widehat{\boldsymbol{\alpha}}_{\beta})\boldsymbol{m}}}(\boldsymbol{m}%
^{T}\widehat{\boldsymbol{\alpha}}_{\beta}-d)>z_{\frac{\alpha}{2}%
}|\boldsymbol{\alpha}=\boldsymbol{\alpha}^{\ast}\right) \nonumber \\
&  =2\Pr\left(  \sqrt{\frac{K}{\boldsymbol{m}^{T}\boldsymbol{\bar{J}}_{\beta
}^{-1}(\widehat{\boldsymbol{\alpha}}_{\beta})\boldsymbol{\bar{K}}_{\beta
}(\widehat{\boldsymbol{\alpha}}_{\beta})\boldsymbol{\bar{J}}_{\beta}%
^{-1}(\widehat{\boldsymbol{\alpha}}_{\beta})\boldsymbol{m}}}\boldsymbol{m}%
^{T}(\widehat{\boldsymbol{\alpha}}_{\beta}-\boldsymbol{\alpha}^{\ast})>\right.
\nonumber \\
&  \hspace*{3cm}\left.  z_{\frac{\alpha}{2}}-\sqrt{\frac{K}{\boldsymbol{m}%
^{T}\boldsymbol{\bar{J}}_{\beta}^{-1}(\widehat{\boldsymbol{\alpha}}_{\beta
})\boldsymbol{\bar{K}}_{\beta}(\widehat{\boldsymbol{\alpha}}_{\beta
})\boldsymbol{\bar{J}}_{\beta}^{-1}(\widehat{\boldsymbol{\alpha}}_{\beta
})\boldsymbol{m}}}(\boldsymbol{m}^{T}\boldsymbol{\alpha}^{\ast}-d)\right)  .
\label{eq:reuslt10}
\end{align}

Finally, since  $\boldsymbol{m}^{T}\boldsymbol{\bar{J}}_{\beta}%
^{-1}(\widehat{\boldsymbol{\alpha}}_{\beta})\boldsymbol{\bar{K}}_{\beta
}(\widehat{\boldsymbol{\alpha}}_{\beta})\boldsymbol{\bar{J}}_{\beta}%
^{-1}(\widehat{\boldsymbol{\alpha}}_{\beta})\boldsymbol{m}$ is a consistent
estimator of $\boldsymbol{m}^{T}\boldsymbol{\bar{J}}_{\beta}^{-1}%
(\boldsymbol{\alpha}^{\ast})\boldsymbol{\bar{K}}_{\beta}(\boldsymbol{\alpha
}^{\ast})\boldsymbol{\bar{J}}_{\beta}^{-1}(\boldsymbol{\alpha}^{\ast
})\boldsymbol{m}$ and
\begin{small}
\[
\boldsymbol{m}^{T}\sqrt{K}(\widehat{\boldsymbol{\alpha}}_{\beta}%
-\boldsymbol{\alpha}^{\ast})\overset{\mathcal{L}%
}{\underset{K\mathcal{\rightarrow}\infty}{\longrightarrow}}\mathcal{N}%
(0,\boldsymbol{m}^{T}\boldsymbol{\bar{J}}_{\beta}^{-1}(\boldsymbol{\alpha
}^{\ast})\boldsymbol{\bar{K}}_{\beta}(\boldsymbol{\alpha}^{\ast}%
)\boldsymbol{\bar{J}}_{\beta}^{-1}(\boldsymbol{\alpha}^{\ast})\boldsymbol{m}%
),
\]
\end{small}
the desired result follows.


\begin{thebibliography}{99}                                                                                               

\bibitem {Basu1} Basu, A., Harris, I. R., Hjort, N. L. and Jones, M. C. (1998).
Robust and efficient estimation by minimizing a density power divergence.
\emph{Biometrika}, \textbf{85}, 549--559.

\bibitem {Basu2} Basu, A., Shioya, H. and Park, C. (2011). \emph{Statistical
Inference: The Minimum Distance Approach}. Chapman $\&$ Hall/CRC Press, Boca  Raton, Florida.

\bibitem {Basu3} Basu, A., Mandal, A., Martin, N. and Pardo, L. (2016).
Generalized Wald-type tests based on minimum density power divergence
estimators. \emph{Statistics}, \textbf{50}, 1--26.

\bibitem {Bala1}Balakrishnan, N. and Ling, M. H. (2012). EM algorithm for
one-shot device testing under the exponential distribution.
\emph{Computational Statistics \& Data Analysis}, \textbf{56}, 502--509.

\bibitem {Bala2}Balakrishnan, N. So, H.Y. and Ling, M. H. (2016a). A Bayesian approach for one-shot device testing with exponential lifetimes under competing risks.
\emph{IEEE Transactions on Reliability}, \textbf{65}, 469--485.


\bibitem {Bala3}Balakrishnan, N. So, H.Y. and Ling, M. H. (2016b). Likelihood inference under proportional hazards model for one-shot device testing. \emph{IEEE Transactions on Reliability}, \textbf{65}, 446--458.

\bibitem {Chimi1}Chimitova, E. V. and Balakrishnan, N. (2015). Goodness-of-fit
tests for one-shot device testing data. In \emph{Advanced Mathematical and
Computational Tools in Metrology and Testing X}. Edited by F. Pavese, W.
Bremser, A. Chunovkina, N. Fischer and A. B. Forbe, World Scientific
Publishing, pp. 124--133, Singapure.


\bibitem {Fan1}Fan, T. H., Balakrishnan, N. and Chang, C. C. (2009). Bayesian
approach for highly reliable electro-explosive devices using one-shot device
testing. \emph{Journal of Statistical Computation and Simulation},
\textbf{79}, 1143--1154.

\bibitem {Ghosh1}Ghosh, A. and Basu, A. (2013). Robust estimation for independent
non-homogeneous observations nusing power density divergence with applications
to linear regression. \emph{Electronic Journal of Statistics, }\textbf{7},
240--2456\emph{. }

\bibitem {Ghosh2}Ghosh, A., Mandal, A., Mart\'{\i}n, N. and Pardo, L. (2016).
Influence analysis of robust Wald-type tests. \emph{Journal of Multivariate
Analysis,} \textbf{147}, 102--126.
 
\bibitem {Lind1} Lindsey, J. and  Ryan, L. (1993). A three-state multiplicative model for rodent tumorigenicity experiments \emph{Journal of the Royal Statistical Society, Series C (Applied Statistics)}, \textbf{42}, 283--300.

\bibitem {Pardo1} Pardo, L. (2005).\emph{ Statistical Inference Based on Divergence
Measures. Statistics: }Chapman \& Hall/CRC Press, New York.

\bibitem {Rodri1}Rodrigues, J., Bolfarine, H. and Louzada-Nieto, F. (1993).
Comparing several accelerated life methods. \emph{Communications in Statistics
Theory and Methods,} \textbf{22}, 2297--2308.




\end{thebibliography}
\end{document}